\documentclass[aps,prc,preprint,superscriptaddress,showkeys]{revtex4-2}
\usepackage{graphicx}
\usepackage{dcolumn}
\usepackage{bm}
\usepackage{amsmath}
\begin{document}

\title{Many-channel microscopic cluster model of $^{8}$Be. \\I. Formation of high-energy resonance states}

\author{V. I. Zhaba}
\email{viktorzh@meta.ua}
\affiliation{Bogolyubov Institute for Theoretical Physics, 03143 Kyiv, Ukraine}
\author{Yu. A. Lashko}
\email{ylashko@gmail.com}
\affiliation{Bogolyubov Institute for Theoretical Physics, 03143 Kyiv, Ukraine}
\affiliation{National Institute for Nuclear Physics, Padova Division, 35131 Padova, Italy}
\author{V. S. Vasilevsky}
\email{vsvasilevsky@gmail.com}
\affiliation{Bogolyubov Institute for Theoretical Physics, 03143 Kyiv, Ukraine}
\date{\today}

\begin{abstract}
The nature and structure of high-energy resonance states in $^{8}$Be, located just below and above the $p+^{7}$Li threshold, are investigated in detail. A microscopic many-cluster and many-channel model is employed to study the formation of these resonances. This model includes three distinct three-cluster configurations: $^{4}$He+$^{3}$H+$p$, $^{4}$He+$^{3}$He+$n$, and $^{4}$He+$d$+$d$, enabling a comprehensive treatment of all major binary decay channels of $^{8}$Be, namely $^{4}$He+$^{4}$He, $p+^{7}$Li, $n+^{7}$Be, and $d+^{6}$Li. The primary focus of our analysis is the structure and dominant decay channels of the twin $1^{+}$, $2^{+}$, $3^{+}$, and $4^{+}$ resonance states.
Additionally, we propose and implement a model to clarify how the $2^{+}$ resonance states lying below the $p+^{7}$Li threshold are formed. We demonstrate that these resonances are Feshbach-type states arising due to coupling of the open $^{4}$He+$^{4}$He channel with the closed channels $p+^{7}$Li, $n+^{7}$Be, and $d+^{6}$Li at these energies.
Overall, the present approach provides a realistic description of the experimentally observed resonance spectrum near the $^7$Li+$p$ decay threshold, including negative-parity states $1^-$ and $2^-$. Our results are consistent with other microscopic calculations but offer more detailed insight into the internal structure and decay pathways of these resonances.
\end{abstract}

\maketitle

\section{Introduction}
The nucleus $^{8}$Be, which has no bound states but a rich spectrum of resonances, has long served as a benchmark for microscopic and semi-microscopic models. It has also been the subject of numerous experimental investigations~\cite{1972PhRvL..29.1331B,1994JPhG...20.1973A,2005PhRvC..72e4312P,2005PhRvL..94l2502D,2008JPhG...35a4004C,2008JPhG...35l5108F,2012JPhG...39a5201W,2017ApJ...850..175L} focused on its resonance structure and on various nuclear reactions in which $^{8}$Be appears as a compound system.

Most previous studies (see, for example, Refs. \cite{2014PhRvC..90b4321P,2015Natur.528..111E,PhysRevC.106.034610,2022JPhG...49e5102F,2023UkrJPh..68..3K}) have focused on its low-energy spectrum, particularly the well-known $0^+$, $2^+$, and $4^+$ resonances, which lie above the $^4$He+$^4$He threshold and are widely recognized as alpha-cluster states. However, the high-energy resonance states, with excitation energies above 16 MeV near the $p+^{7}$Li threshold, have received significantly less attention. These states exhibit a more complex structure and cannot be described solely within the conventional alpha-cluster picture.  

A particularly intriguing feature of $^8$Be is the presence of twin resonances — pairs of resonance states with the same total angular momentum and parity but slightly different energies. Such twin states have been observed for the $2^+$, $1^+$, and $3^+$ resonances, with another pair yet to be assigned definitive quantum numbers. The origin and nature of these twin resonances remain open questions, motivating the present study.  

Furthermore, the level structure of $^8$Be near the $^7$Li + $p$ threshold is of considerable interest in both nuclear physics and astrophysics, yet it remains only partially understood. Although narrow negative-parity states are absent, the broad $1^-$ and $2^-$ states remain the subject of debate. Understanding the interplay between these structures, the role of cluster polarization, and the effects of interchannel coupling is crucial for developing a more complete theoretical framework. This work aims to provide new insights into the high-energy spectrum of $^8$Be using a microscopic three-cluster approach, with a particular focus on the twin resonance phenomenon and its underlying dynamics.

High-energy states of $^8$Be near the $^7$Li$+p$ decay threshold have been investigated using various theoretical approaches, including the Generator Coordinate Method \cite{1994NuPhA.573...28D}, the Complex Scaling Method Refs. \cite{2020PTEP.2020lA101M},
\cite{2014PrPNP..79....1M}, the Gamow Shell Model \cite{Fernandez2024}, the ab initio no-core shell model \cite{PhysRevC.87.014327}, the no-core shell model with continuum \cite{2024PhRvC.110a5503G}, Green’s function Monte Carlo calculations \cite{PhysRevC.88.044333}, the tensor-optimized shell model \cite{10.1093/ptep/ptu012}, variational Monte Carlo calculations \cite{PhysRevC.62.014001}, and the Fermionic Molecular Dynamics model \cite{Henninger_2016}. However, only a few models have estimated not only the energies but also the widths of these states. In this work, we will focus on reviewing only those studies in more detail, as we aim to compare our results with theirs.

Among these, the NCSMC approach has provided a valuable framework for investigating both structure and reactions in $^8$Be, explicitly including continuum effects and realistic nucleon-nucleon interactions. A recent study by Navrátil et al. \cite{2024PhRvC.110a5503G} applied this method to analyze the high-energy spectrum of $^8$Be, focusing on the key $1^+$ resonances near the $^7$Li$+p$ threshold. While their calculations successfully captured the overall structure of these states, their positions and widths did not fully align with experimental values without additional phenomenological corrections. This highlights the ongoing challenges in achieving fully predictive \textit{ab initio} descriptions of near-threshold nuclear resonances. In the present work, we take a complementary approach by employing a many-cluster, many-channel model, aiming to provide an alternative perspective on the resonance structure and decay dynamics in this energy region.

In~\cite{1994NuPhA.573...28D}, Descouvemont and Baye employed a microscopic three-cluster model within the Generator Coordinate Method (GCM) to study the spectroscopy of $^8$Be and its role in $^7$Li+p and $^7$Be+n reactions. The model included $\alpha+\alpha$, $\alpha+t+p$, and $\alpha+^3\text{He}+n$ configurations, considering both $^7\text{Li}+p$ and $^5\text{Li}+t$ coupling schemes. In the $\alpha+\alpha$ configuration, Slater determinants were constructed using $\alpha$-cluster wave functions from the harmonic oscillator model with a common oscillator parameter \( b = 1.50 \, \text{fm} \). For the $\alpha+t$ and $\alpha+^3\text{He}$ configurations, a single generator parameter was chosen to minimize the ground-state binding energies of $^7$Li and $^7$Be. The $3/2^-$ ground state, as well as the $1/2^-$, $7/2^-$, and $5/2^-$ excited states of $^7$Li and $^7$Be, were included. For $\alpha+p$ and $\alpha+n$, the $3/2^-$ and $1/2^-$ partial waves were considered. Distortion effects in the seven- and five-nucleon subsystems were neglected.

A single Hamiltonian incorporating the Volkov nucleon-nucleon potential, spin-orbit interaction, and Coulomb force was employed, with key parameters adjusted to reproduce experimental data, including the $^8$Be($2_2^+$) binding energy relative to the $^7$Li+p threshold and the excitation energy of the $1/2^-$ state in $^7$Li.

The GCM successfully described the $^8$Be positive-parity spectrum, including isospin-mixed $2^+$, $1^+$, and $3^+$ doublets, while also predicting a $4^+;T=1$ state near 5.49 MeV  and  the $4^+;T=0$ resonance at 5.99 MeV above the $^7\text{Li}+p$ threshold.  The model also suggested that subthreshold $2_2^+$ and $2_3^+$ states could contribute to the low-energy enhancement of the S-factor observed in $^7$Li(p,$\alpha$)$\alpha$ experiments \cite{engstler1992isotopic}.

For negative-parity states, the model identified broad $1^-$ and $2^-$ resonances near 2.5 MeV and 2 MeV above the proton threshold, respectively, but overestimated the isospin mixing in the $2^-$ state. The calculated $^7$Be(n,p)$^7$Li cross section underestimated experimental data \cite{koehler19887} by a factor of 100, highlighting the dominant experimental role of the $2^-$ resonance, which was not accurately captured by the model.

The paper by Fernández et al. \cite{Fernandez2024} investigates the phenomenon of clustering in nuclear systems, with a particular focus on $^8\text{Be}$, using a microscopic cluster approach based on the Gamow Shell Model (GSM) for open quantum systems. The authors present detailed calculations of the resonance spectrum of $^8\text{Be}$, providing the energies and widths of its positive parity resonant states, including those above the $^7\text{Li} + p$ threshold.

Using the coupled-channel GSM (GSM-CC) formalism, the study describes both the structural and the reaction properties of $^8\text{Be}$. The model accounts for the coupling to various reaction channels, such as $^4\text{He} + ^4\text{He}$, $^7\text{Li} + p$, $^7\text{Be} + n$, and $^6\text{Li} + d$, and treats $^8\text{Be}$ as a system composed of an inert $^4\text{He}$ core and several valence nucleons. The subsystems $^{6,7}\text{Li}$ and $^7\text{Be}$ are similarly modeled with a $^4\text{He}$ core and valence nucleons. An N$^3$LO realistic interaction without a three-body contribution was employed as the nucleon-nucleon interaction. This interaction underbinds $^4\text{He}$ and the deuteron by 2 MeV and 0.25 MeV, respectively. To ensure the correct energy for the $^4$He$+^4\text{He}$ threshold, the authors used the experimental binding energy of $^4\text{He}$ in their calculations.

In \cite{Fernandez2024}, the authors show that coupling to the deuteron channel improves the theoretical description of the spectrum, achieving better agreement with experimental data, particularly for doublets such as $(1^+_1, 1^+_2)$, $(2^+_1, 2^+_2)$, and $(3^+_1, 3^+_2)$.
The authors also emphasize the role of near-threshold states, which reveal emergent cluster structures influenced by interactions with the continuum. These structures are attributed to the collective rearrangement of wave functions driven by their coupling to the environment of open channels. The paper underscores how the inclusion of these effects enables a more accurate description of both the resonance spectrum and the resonance widths, highlighting the importance of the interplay between nuclear structure and reaction dynamics.

Compared to Ref.~\cite{1994NuPhA.573...28D}, our study includes two additional three-cluster configurations of $^8$Be: $\alpha+d+d$ and $\alpha+2p+2n$. This extension, in particular, allowed us to consider the $^6$Li$+d$ binary channel, which was shown in Ref.~\cite{Fernandez2024} to be crucial for describing the $1^+$, $2^+$, and $3^+$ doublets. While we did not include the $5/2^-$ and $7/2^-$ excited states of $^7$Li and $^7$Be, our approach incorporated more realistic wave functions for the binary subsystems ($^7$Li, $^6$Li, $^5$Li, $^5$He, and $^7$Be) and accounted for polarization effects, which, as we show in this paper, are significant for describing the high-energy states of $^8$Be. Furthermore, we employed a rigorous treatment of scattering boundary conditions, enabling us to accurately describe broad resonances, such as the $2^-$ and $1^-$ states, and to extract comprehensive information about these resonances from the energy dependence of phase shifts and inelasticity coefficients.

Our study expands upon the work in Ref.~\cite{Fernandez2024} by including additional binary channels, such as $^5$He+$^3$He and $^5$Li+$^3$H. Moreover, we analyzed not only positive-parity resonance states but also negative-parity states, providing a detailed investigation of the partial widths of the resonances. Additionally, we performed a comprehensive phase-shift analysis of the states above the $^7$Li$+p$ decay threshold of $^8$Be.

This paper constitutes the first part of a two-part study. In a forthcoming second paper (Part II), we will apply the present many-channel microscopic cluster model to explore the dynamics of reactions induced by protons on $^{7}$Li, neutrons on $^{7}$Be, and deuterons on $^{6}$Li. Particular emphasis will be placed on how the resonance states identified in the present work affect the corresponding reaction cross sections and astrophysical $S$-factors. These investigations will provide a more complete understanding of low-energy reaction mechanisms and may offer new insights into the longstanding cosmological lithium problem.

This paper is organized as follows. In Section \ref{Sec:Method}, we describe the microscopic many-cluster, many-channel model of $^8$Be and discuss the choice of model space, nucleon-nucleon interaction, and parameter adjustments used in the present calculations. Section \ref{Sec:Results} presents the analysis of the obtained results. The conclusions are summarized in Section \ref{Sec:Concl}.

\section{Microscopic Many-Channel Cluster Model and Input Parameters} 
\label{Sec:Method}

\subsection{Microscopic Many-Channel Cluster Model}
Here we provide a brief introduction to the method used, as it is explained in detail in Refs. \cite{2009NuPhA.824...37V}, \cite{2017NuPhA.958...78L}, \cite{2024PhRvC.109d5803L}, \cite{2024PhRvC.110c5806L}. 

Starting with a three-cluster configuration $A = A_{1} + A_{2} + A_{3}$, it is natural to approximate the wave function of the compound system in the form:
\begin{equation}
\Psi^{J}=\sum_{\alpha=1}^{3}\widehat{\mathcal{A}}\left\{  \left[  \Phi
_{1}\left(  A_{1},S_{1}\right)  \Phi_{2}\left(  A_{2},S_{2}\right)  \Phi
_{3}\left(  A_{3},S_{3}\right)  \right]  _{S}f_{c,L}^{\left(  \alpha\right)
}\left(  \mathbf{x}_{\alpha},\mathbf{y}_{\alpha}\right)  \right\}
_{J},\label{eq:M001}%
\end{equation}
where $\widehat{\mathcal{A}}$ is the antisymmetrization operator, $\Phi_{\alpha}\left(  A_{\alpha},S_{\alpha}\right)  $ is the shell-model
wave function describing the internal motion of $A_{\alpha}$ nucleons within the
cluster $\alpha$, with $S_{\alpha}$ being the spin of cluster $\alpha$, and $\ f_{\alpha}\left(  \mathbf{x}_{\alpha
},\mathbf{y}_{\alpha}\right)  $ is a Faddeev component that describes the relative
motion of clusters. The total spin of the system is denoted as $S$, the total orbital momentum as $L$, and the total angular momentum as $J.$ 
The vector $\mathbf{x}_{\alpha}$ is the Jacobi vector, proportional to the distance between clusters $\beta$ and $\gamma$, while
$\mathbf{y}_{\alpha}$ is a Jacobi vector connecting cluster $\alpha$ to the center of mass of clusters $\beta$ and $\gamma$. The indices
$\alpha$, $\beta$, and $\gamma$ form a cyclic permutation of 1, 2, and 3. 

In Eq.~(\ref{eq:M001}), the sum over index $\alpha$ implies that the selected three-cluster configuration includes three binary channels, $A_{\alpha} + \left( A_{\beta} + A_{\gamma} \right) $, each of which describes the scattering of the cluster with index $\alpha$ on a bound or pseudo-bound state formed by clusters $\beta$ and $\gamma$. Note that in our model, pseudo-bound states represent resonance states or specific states of a two-cluster continuum.

In the present model, we neglect the three-cluster continuum and restrict ourselves to a set (as large as possible) of binary channels, where one of the components of each binary channel is treated as a two-cluster subsystem. Thus, with
some minor simplifications, we reduce the three-cluster wave function
(\ref{eq:M001}) to
\begin{equation}
\Psi^{\left(  E,J\right)  }=\sum_{\alpha}\widehat{\mathcal{A}}\left\{
\Phi_{\alpha}\left(  A_{\alpha},S_{\alpha}\right)  \psi_{\alpha}\left(
\mathcal{E}_{\alpha},A_{\beta}+A_{\gamma},j_{2}\right)  \varphi_{E-\mathcal{E}%
_{\alpha},l_{\alpha},j_{1}}\left(  \mathbf{y}_{\alpha}\right)  \right\}
_{J},\label{eq:M004}%
\end{equation}
where $l_{\alpha}$ and $j_{1}$ are the orbital and angular momenta of the
relative motion of cluster $\alpha$ around the two-cluster $\beta+\gamma$
subsystem, and
\begin{equation}
\psi_{\alpha}\left(  \mathcal{E}_{\alpha},A_{\beta}+A_{\gamma},j_{2}\right)
=\widehat{\mathcal{A}}\left\{  \left[  \Phi_{\beta}(A_{\beta},S_{\beta}%
)\Phi_{\gamma}(A_{\gamma},S_{\gamma})\right]  _{S_{\beta\gamma}}%
\chi_{\mathcal{E}_{\alpha},\lambda_{\alpha}}(\mathbf{x}_{\alpha})\right\}
_{j_{2}}\label{eq:M005}%
\end{equation}
is the wave function of the two-cluster subsystem with energy $\mathcal{E}_{\alpha}$,
two-cluster spin $S_{\beta\gamma}$, orbital momentum $\lambda_{\alpha}$, and
angular momentum $j_{2}$.

Here, $\varphi_{E-\mathcal{E}_{\alpha},l_{\alpha},j_{1}}(\mathbf{y}_{\alpha})$ describes the relative motion of cluster $\alpha$ with respect to the center of mass of the two-cluster subsystem $\beta+\gamma$, while $\chi_{\mathcal{E}_{\alpha},\lambda_{\alpha}}(\mathbf{x}_{\alpha})$ represents the internal wave function of the two-cluster subsystem with energy $\mathcal{E}_{\alpha}$ and orbital momentum $\lambda_{\alpha}$.

The three-cluster wave function in Eq.~(\ref{eq:M001}) is expressed in the $LS$ coupling scheme, while the wave function in Eq.~(\ref{eq:M004}) is given in the $JJ$ coupling scheme. The latter more adequately reflects physical reality and is used for the calculation of elements of the scattering $S$-matrix, whereas the $LS$ coupling scheme is simpler for obtaining analytical expressions for matrix elements of the norm kernel and Hamiltonian, and for performing their numerical calculations. After calculating the matrix elements of all necessary operators in the $LS$ coupling scheme, we transform them to the $JJ$ coupling scheme, as explained in Section~II.B of Ref.~\cite{2024PhRvC.109d5803L}.

An important feature of the present model is its ability to account for the polarization of the two-cluster subsystem as it approaches the third cluster. Here, polarization refers to the ability of the two-cluster subsystem to modify its internal structure — such as its size or shape—due to the interaction with the third cluster. To this end, the wave function $\varphi_{E-\mathcal{E}_{\alpha},l_{\alpha},j_{1}}(\mathbf{y}_{\alpha})$ in Eq.~(\ref{eq:M004}) is expanded in a basis of oscillator functions, which also facilitates the imposition of proper boundary conditions. The internal wave function of the two-cluster subsystem, $\chi_{\mathcal{E}_{\alpha},\lambda_{\alpha}}(\mathbf{x}_{\alpha})$ in Eq.~(\ref{eq:M005}), is expanded in a basis of Gaussian functions.

Solving the Schrödinger equation with $N_G$ Gaussian basis functions yields $N_G$ eigenvalues $\mathcal{E}_{\alpha,\sigma}$ ($\sigma = 1, 2, \ldots, N_G$) and corresponding eigenfunctions. The lowest eigenvalue $\mathcal{E}_{\alpha,1}$ typically corresponds to the ground state of the two-cluster subsystem, while the remaining eigenvalues represent pseudo-bound states—discretized approximations to the continuum obtained by diagonalization in a finite basis.

When only the eigenfunction associated with $\mathcal{E}_{\alpha,1}$ is retained in Eq.~(\ref{eq:M004}), the two-cluster subsystem has a fixed size and polarization effects are excluded. In contrast, including multiple eigenfunctions ($N_\sigma = 2, 3, \ldots, N_G$) allows the subsystem to adjust its internal structure dynamically, thereby incorporating polarization effects.

In what follows, calculations using only the ground-state eigenfunction are labeled “No” (no polarization), whereas those including additional eigenfunctions are labeled “Yes” (with polarization). Further details on the treatment of cluster polarization in this approach can be found in Refs.~\cite{2009NuPhA.824...37V, 2017NuPhA.958...78L}.

It should be noted that symmetry restrictions imposed by the $\alpha+\alpha$ channel have a significant impact on the resonance structure of $^{8}$Be and on the cross sections of various reactions. The $\alpha+\alpha$ system can be treated as a two-boson system, since the spin of the alpha particle is zero. As a result, the total wave function of the $\alpha+\alpha$ system must be symmetric under particle interchange, which implies that only even-parity states with even values of the total angular momentum $J$ are allowed.

These symmetry conditions impose specific constraints on the resonance spectrum of $^{8}$Be. First, only even-parity states with even $J$ can appear in the energy region between the $\alpha+\alpha$ and $p+^{7}$Li thresholds, where the $\alpha+\alpha$ channel is the only open channel. Second, above the $p+^{7}$Li threshold, the $\alpha+\alpha$ channel continues to contribute to the formation of resonance states, but only those with positive parity and even total angular momentum. In addition, the $\alpha+\alpha$ channel may serve as either the entrance or exit channel in reactions, but only in states with even $J$ and positive parity.

\subsection{Model Space and Interaction Parameters} \label{Sec:Input}

Since the primary focus of our paper is the analysis of the excited states near the $^7$Li+p decay threshold of $^8$Be, it is essential to consider all three-cluster configurations that can generate the most significant binary channels in this energy range. To achieve this, we employ two distinct model spaces, each designed to account for a specific set of three-cluster configurations and the corresponding binary channels.
The first model space consists of three three-cluster configurations:
\begin{equation}
^{4}\text{He}+^{3}\text{H}+p,\quad^{4}\text{He}+^{3}\text{He}+n,\quad
^{4}\text{He}+d+d,\quad\label{eq:N001}%
\end{equation}
which allow us to account for the following binary channels:
\begin{equation}
p+^{7}\text{Li},\quad ^{3}\text{H}+^{5}\text{Li},\quad n+^{7}\text{Be},\quad ^{3}\text{He}+^{5}\text{He},\quad 
d+^{6}\text{Li},\quad ^{4}\text{He}+^{4}\text{He}. \label{eq:N002}%
\end{equation}
The second model space extends the first by including an additional three-cluster configuration:
\begin{equation}
^{4}\text{He}+^{2}p+^{2}n,\label{eq:N003}%
\end{equation}
which enables us to consider two additional binary channels:
\begin{equation}
^{2}n+^{6}\text{Be},\quad ^{2}p+^{6}\text{He}. \label{eq:N004}%
\end{equation}

It is noteworthy that both new channels make a significant contribution to the total isospin of the compound system. Specifically, the isospin contributions are as follows: $T=2$ contributes $1/6$, $T=1$ contributes $1/2$ (the largest contribution), and $T=0$ contributes $1/3$. Thus, it is expected that the configuration (\ref{eq:N003}) or the binary channels (\ref{eq:N004}) will enhance resonance states with a total isospin $T=1$. Below, we will show that these channels are important for describing the $2^+$ isospin doublet resonances.

In what follows, we will refer to the first model space as the six-channel space (6Chs) and the second model space as the eight-channel space (8Chs). Note that the additional channels, which have a total spin value of $S=0$, can contribute only to states with normal parity $\pi=\left(  -1\right)  ^{J}$, where $J$ is the total angular momentum of $^{8}$Be.

The first two channels, $p+^{7}$Li and $^{3}$H$+^{5}$Li, are generated by the three-cluster configuration $^{4}$He+$^{3}$H+$p$, the next two channels, $n+^{7}$Be and $^{3}\mathrm{He}+^{5}\mathrm{He},$ are associated with the $^{4}$He+$^{3}$He+$n$ configuration, and the fifth binary channel, $^{6}$Li+$d$, is formed by the $^{4}$He+$d+d$ configuration. The last binary channel, $^{4}$He+$^{4}$He, can be generated by all four configurations shown in Eqs.~(\ref{eq:N001}) and (\ref{eq:N004}).

We have chosen the Hasegawa-Nagata (HNP) potential as the nucleon-nucleon interaction.
Two criteria are employed to select the input parameters of the model: the oscillator length $b$ and the parameters of the chosen nucleon-nucleon potential. The first criterion is that the oscillator length $b=1.362$ fm is chosen to minimize the threshold energy of the lowest three-cluster configuration, $^{4}$He+$^{3}$H+$p$. The second criterion is that the Majorana parameter $m=-0.008$ of the HNP, along with the spin-orbit strength  $f_{LS}=0.3$, are adjusted to reproduce the energies of the ground 3/2$^{-}$ and first excited 1/2$^{-}$ states of $^{7}$Li and $^{7}$Be. Table~\ref{Tab:InputParams} lists the energies of these states obtained with the selected parameters.

\begin{table}[htbp]
\centering
\caption{
Energies \( \mathcal{E}(j_2^\pi) \) (in MeV) and rms mass radii \( R_m(j_2^\pi) \) (in fm) of the bound states in the \( ^7\mathrm{Li} \) and \( ^7\mathrm{Be} \) two-cluster subsystems. The theoretical results are obtained using the Hasegawa–Nagata NN potential with Majorana parameter \( m = -0.008 \), spin-orbit strength \( f_{LS} = 0.30 \), and oscillator length \( b = 1.362 \) fm. Experimental values of the energies are taken from Ref.~\cite{2002NuPhA.708....3T}, and those of the rms mass radii from Ref.~\cite{TANIHATA2013215}.}
\vspace{0.5em}
\begin{tabular}{|c|c|c|c|c|c|}
\hline
Nucleus &  & \( \mathcal{E}\left( \frac{3}{2}^{-} \right) \) & \( \mathcal{E}\left( \frac{1}{2}^{-} \right) \) & \( R_{m}\left( \frac{3}{2}^{-} \right) \) & \( R_{m}\left( \frac{1}{2}^{-} \right) \) \\
\hline
\( ^7 \)Li 
& Theory & -2.447 & -1.974 & 2.29 & 2.35 \\
\cline{2-6}
& Exp & -2.467 & -1.989 & 2.33(5) & – \\
\hline
\( ^7 \)Be 
& Theory & -1.597 & -1.150 & 2.34 & 2.42 \\
\cline{2-6}
& Exp & -1.587 & -1.158 &  2.31(2) & - \\
\hline
\end{tabular}
\label{Tab:InputParams}
\end{table}

\begin{figure}[ptbh]
\begin{center}
\includegraphics[height=3.9617in,width=5.0868in]{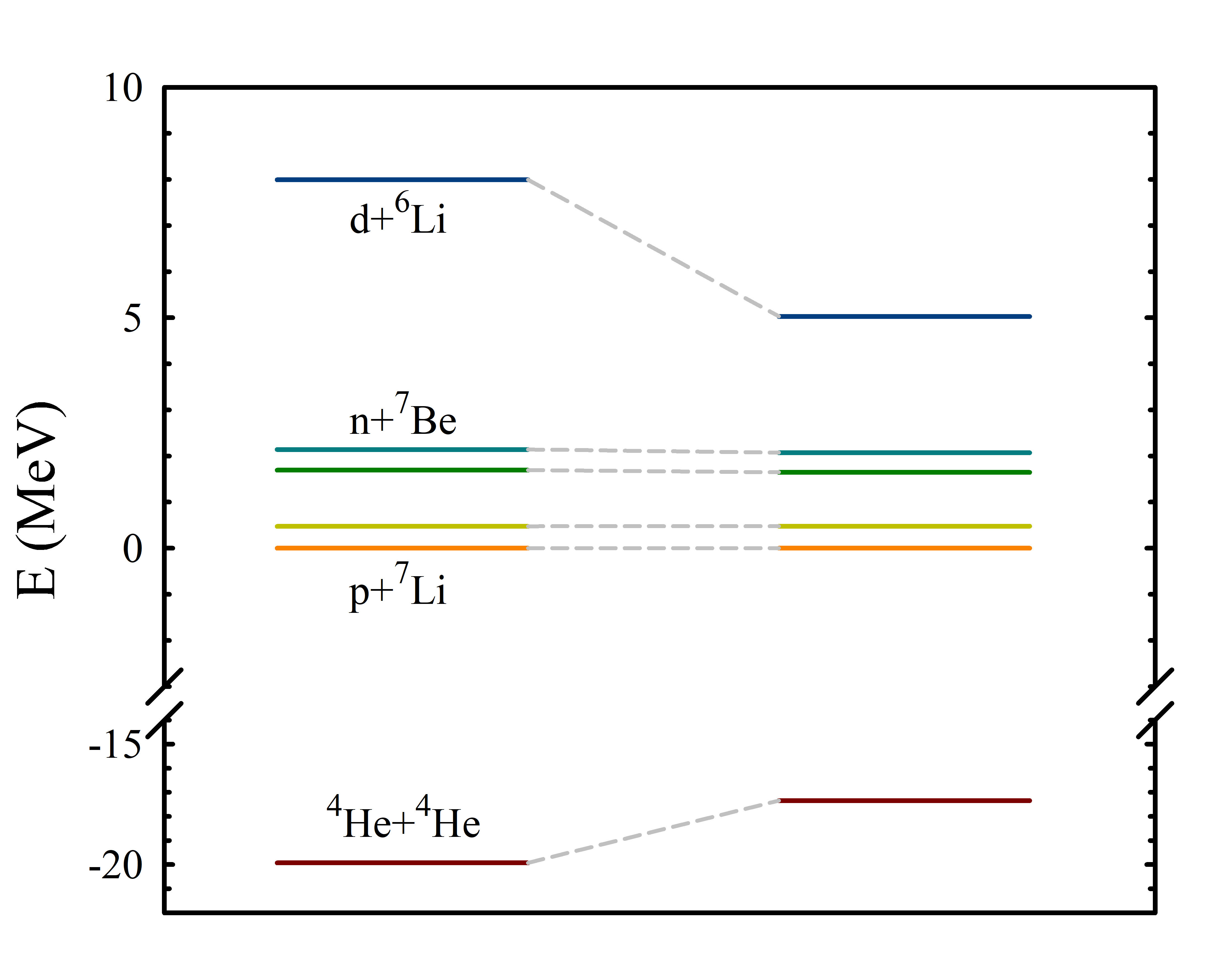}
\caption{Threshold energies for $^{8}$Be decay into binary channels. The left panel shows the theoretical values, while the right panel displays the experimental values \cite{2004NuPhA.745..155T}.}
\label{Fig:ThresholdsHNP}
\end{center}
\end{figure}

The relative positions of the main binary thresholds of $^{8}$Be are shown in Fig.~\ref{Fig:ThresholdsHNP}. The threshold energy is measured relative to the $p+^{7}$Li threshold. This channel, along with the $n+^{7}$Be channel, is represented by two threshold energies: one corresponding to the ground 3/2$^{-}$ state of $^{7}$Li or $^{7}$Be, and the other corresponding to the first excited 1/2$^{-}$ state of these nuclei.
The present model with the selected input parameters correctly reproduces the relative position of the $p+^{7}$Li and $n+^{7}$Be thresholds. However, the relative positions of the $^{4}$He+$^{4}$He and $d+^{6}$Li thresholds with respect to the $p+^{7}$Li threshold differ by 2.6 MeV and 3.0 MeV, respectively.
Notably, this discrepancy is less pronounced than the energy difference between the $^7$Li+$p$ and $\alpha+\alpha$ thresholds reported in \cite{1994NuPhA.573...28D}, which can be attributed to the use of the Hasegawa-Nagata potential instead of the Volkov potential for the nucleon-nucleon interaction.

\section{Results \label{Sec:Results}}

\subsection{Resonance parameters of twin states}
\label{sec:resonance_parameters}

In Table~\ref{Tab:2PRHNPvsModel}, we present the parameters of the 2$^{+}$ resonance states found near the $p+^{7}$Li threshold using the HNP potential.  
To investigate the nature of these states, we employ both six- and eight-channel approximations and consider cases with and without polarization of the two-cluster subsystems.

The results demonstrate that cluster polarization plays a significant role in the formation of the $2^{+}$ twin resonance states, with the most pronounced effect seen in their energies. Specifically, polarization lowers the energy of the first resonance state from 0.014 MeV to $-0.880$ MeV in the 6Chs case (from $-0.002$ MeV to $-0.816$ MeV in the 8Chs case). It also shifts the energy of the second resonance state downward by approximately 0.8 MeV and 1.1 MeV in the 6Chs and 8Chs cases, respectively.

The parameters obtained with eight channels and polarization are in the best agreement with experimental data. The inclusion of the $^6$He+$^2p$ and $^6$Be+$^2n$ channels reduces the energy splitting between the two twin $2^+$ resonances and, consequently, lowers the second $2^+$ state below the $^7$Li+$p$ decay threshold. As a result, the energy of this resonance closely matches the experimental energy of $-0.262$ MeV, although the first $2^+$ resonance remains approximately 190 keV below the corresponding experimental energy.

\begin{table}[tbp] \centering
\caption{Energies $E$ and widths $\Gamma$ of the $2^{+}$ resonance states obtained in different model approximations using the Hasegawa–Nagata potential. Experimental values from Ref.~\cite{2004NuPhA.745..155T} are shown for comparison.}
\begin{tabular}
[c]{|c|c|c|c|c|c|}\hline
Model & Polarization &  $E$, MeV & $\Gamma$, keV & $E$,
MeV & $\Gamma$, keV\\\hline
6Chs & No &  0.014 & 7.00 & 0.924 & 53.16\\
6Chs & Yes &  -0.880 & 3.50 & 0.126 & 31.99\\\hline
8Chs & No &  -0.002 & 9.59 & 0.818 & 52.82\\
8Chs & Yes &  -0.816 & 16.82 & -0.238 & 243.08\\\hline
Exp & - &  -0.628 & 108. & -0.262 & 74.\\\hline
\end{tabular}
\label{Tab:2PRHNPvsModel}
\end{table}

\begin{figure}[ptb]
\begin{center}
\includegraphics[height=11.0952cm,width=13.641cm]{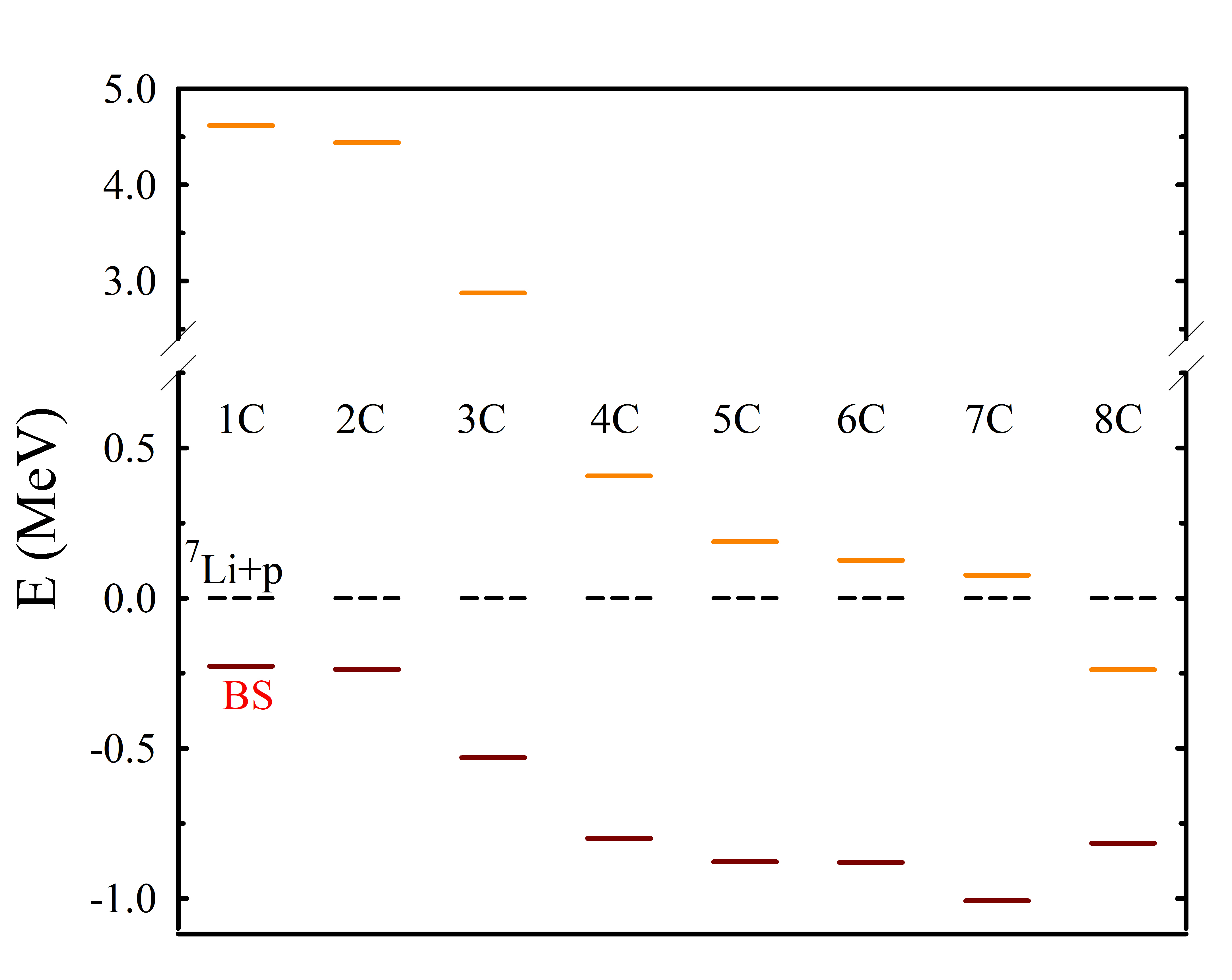}
\caption{Spectrum of the twin $2^{+}$ resonance states obtained with different numbers of binary channels. ``BS'' denotes a bound state in the 1C approximation. The notations 1C through 8C correspond to the sequential addition of the following channels: $p+^{7}$Li, $^{4}$He+$^{4}$He, $^{3}$H+$^{5}$Li, $n+^{7}$Be, $^{3}$He+$^{5}$He, $d+^{6}$Li, $^{2}n+^{6}$Be, and $^{2}p+^{6}$He.}
\label{Fig:Resons2PHNPvsChs}
\end{center}
\end{figure}

Now we consider how the two 2$^{+}$ resonances are formed and which channels play an
important role in their formation.  
In Figure~\ref{Fig:Resons2PHNPvsChs}, we present the energies of the 2$^{+}$ resonance states as a function of the number of binary channels included in the calculations. Initially, a single-channel approximation (1C) involves only the $p+^{7}$Li channel. This approximation generates several bound states, one of which is shown in Figure~\ref{Fig:Resons2PHNPvsChs}, along with a shape resonance at a relatively high energy of $E=4.6$ MeV.
In the 2C approximation, the $p+^{7}$Li channel is coupled to the $^{4}$He+$^{4}$He channel.
Next, the $^{3}$H+$^{5}$Li channel is added (3C), followed by the $n+^{7}$Be channel (4C), and then the $^{3}$He+$^{5}$He channel (5C). Finally, by including the binary channels $d+^{6}$Li, $^{2}n+^{6}$Be, and $^{2}p+^{6}$He, we form the 6C, 7C, and 8C cases, respectively.

The order of inclusion of binary channels in Fig.~\ref{Fig:Resons2PHNPvsChs} is determined by several factors. We begin with the $p+^{7}\text{Li}$ channel because we are investigating the nature of resonance states near the $p+^{7}\text{Li}$ decay threshold of $^8\text{Be}$. This channel plays a primary role in the formation of these resonance states due to the proximity of the corresponding decay threshold. 
The channels $p+^{7}\text{Li}$ and $^{3}\text{H}+^{5}\text{Li}$ arise from the three-cluster configuration $^{4}\text{He}+^{3}\text{H}+p$, while $n+^{7}\text{Be}$ and $^{3}\text{He}+^{5}\text{He}$ originate from $^{4}\text{He}+^{3}\text{He}+n$, and $d+^{6}\text{Li}$ is associated with the $^{4}\text{He}+d+d$ configuration. 
The next criterion is the ordering of channels by threshold energy. Binary channels associated with a given three-cluster configuration are added sequentially, starting with those that have the lowest threshold energy. 
The $^{4}\text{He}+^{4}\text{He}$ channel is included immediately after $p+^{7}\text{Li}$ because it has the lowest decay threshold and can be generated by all of the above-mentioned three-cluster configurations.

Comparing case 1C and case 2C in Fig.~\ref{Fig:Resons2PHNPvsChs}, we observe that the inclusion of the $^{4}$He+$^{4}$He channel slightly shifts the position of the shape resonance and transforms the bound state into a resonance state with approximately the same energy. This indicates that the resulting resonance state can be classified as a Feshbach resonance~\cite{1958AnPhy...5..357F,1962AnPhy..19..287F}, caused by weak coupling between the $p+^{7}$Li and $^{4}$He+$^{4}$He channels.
A qualitative assessment of the coupling strength between the $p+^{7}$Li and $^{4}$He+$^{4}$He channels will be provided in Section~\ref{sec:feshbach}.

Starting from case 3C, the addition of new channels significantly lowers the energies of both the shape and Feshbach resonances. An exception occurs for the Feshbach resonance: when the eighth channel, $^{2}p+^6$He, is added, the energy of the lowest resonance state increases by 0.2 MeV, leading to a decrease in the energy splitting between the shape and Feshbach resonances.

The spectra of the $1^+$ and $3^+$ resonance states, determined with and without cluster polarization, are presented in Table~\ref{Tab:1P3PRHNPvsModel}. The calculations were performed using the 5Ch approximation, as the $^4$He+$^4$He, $^{2}n+^6$Be, and $^{2}p+^6$He channels do not contribute to the formation of the $1^+$ and $3^+$ states in $^8$Be.

Table~\ref{Tab:1P3PRHNPvsModel} demonstrates that cluster polarization has a significant impact on the energies and widths of the twin $1^+$ and $3^+$ resonance states. For example, cluster polarization reduces the energy of the first $1^+$ resonance state by more than a factor of two and decreases its width by a factor of five. The second $1^+$ resonance state, being broader, does not undergo such dramatic modifications: its energy is reduced by about 25\%, while its width increases slightly.

A similar pattern is observed for the $3^+$ resonance states. However, without cluster polarization, only one resonance state with a relatively large width is obtained. When cluster polarization is included, two $3^+$ resonance states emerge, both with lower energies and roughly half the width compared to the resonance state obtained without polarization.

For both the twin $1^+$ and $3^+$ resonance states, our results show good agreement with experimental data. In particular, the energies and widths of the lowest $1^+$ and $3^+$ resonances match the experimental values very well.

\begin{figure}[ptb]
\begin{center}
\includegraphics[height=10.9458cm,width=13.6652cm]{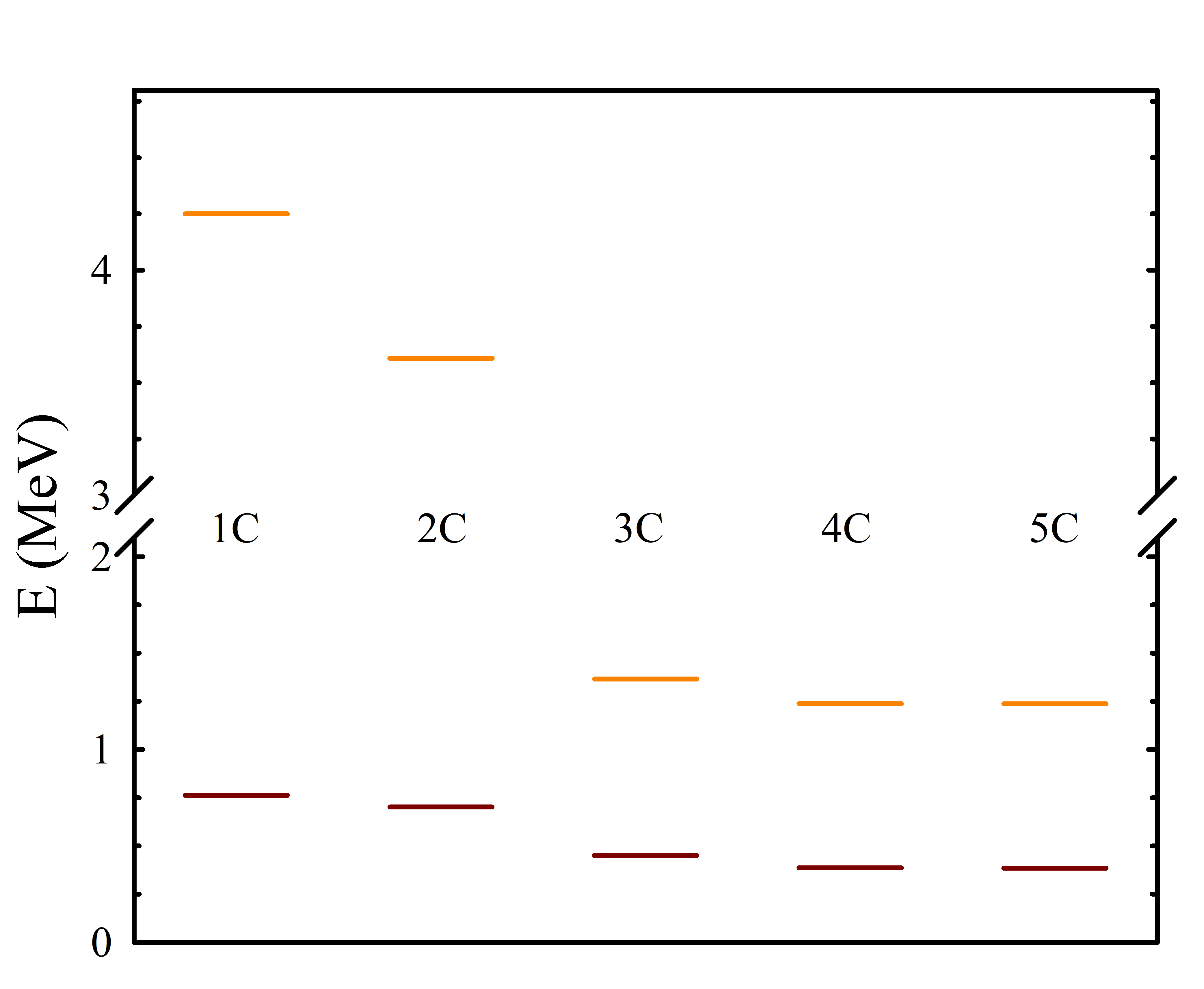}
\caption{Energies of the $1^{+}$ resonance states as a function of the number of binary channels included in the calculations. The labels from 1C to 5C correspond to the sequential addition of the binary channels: $p+^{7}$Li, $^{3}$H+$^{5}$Li, $n+^{7}$Be, $^{3}$He+$^{5}$He, and $d+^{6}$Li.} 
\label{Fig:Resons1PHNPvsChs}
\end{center}
\end{figure}

\begin{table}[tbph] \centering
\caption{Energies ($E$) and widths ($\Gamma$) of the twin $1^+$ and $3^+$ resonance states in $^8$Be, obtained with and without cluster polarization. Experimental values from Ref.~\cite{2004NuPhA.745..155T} are shown for comparison.}
\begin{tabular}
[c]{|c|c|c|c|c|c|c|}\hline
& Model & Polarization &  $E$, MeV & $\Gamma$, keV & $E$, MeV & $\Gamma$, keV\\\hline
1$^{+}$ & 5Chs & No & 0.843 & 47.98 & 1.647 & 287.25\\\cline{2-7}
& 5Chs & Yes  & 0.385 & 9.63 & 1.237 & 350.90\\\cline{2-7}
& Exp & - &  0.386 & 11. & 0.869 & 138.\\\hline\hline
3$^{+}$ & 5Chs & No &  3.344 & 589.52 &  & \\\cline{2-7}
& 5Chs & Yes  & 1.671 & 280.91 & 2.339 & 260.56\\\cline{2-7}
& Exp & - &  1.816 & 270 & 1.986 & 230\\\hline
\end{tabular}
\label{Tab:1P3PRHNPvsModel}
\end{table}
The formation of the 1$^{+}$ resonance states is shown in Fig. \ref{Fig:Resons1PHNPvsChs}. The $p+^{7}$Li channel generates two resonance states, one of which lies at relatively high energies, exceeding 4 MeV.
The inclusion of additional channels does not change the number of resonances but only shifts their energies. Therefore, these resonances are not Feshbach resonances. The magnitude of the energy shift depends on the strength of the coupling between different channels.
The $n+^{7}$Be channel significantly alters the positions of the 1$^{+}$ resonance states, decreasing the energy of the second resonance state by 2.2 MeV. In contrast, the $d+^{6}$Li channel has only a minor effect on the resonance energies.

In Table \ref{Tab:4PRHNPvsModel}, we present the parameters of the 4$^{+}$ resonance
states. According to Ref. \cite{2004NuPhA.745..155T}, only one experimentally observed $4^+$ resonance state exists above the $p+^7$Li threshold, with an energy of $E = 2.606$ MeV and a width of $\Gamma = 700 \pm 100$ keV. However, our model does not predict a resonance state at an energy close to 2.606 MeV. Instead, we obtained two resonance states with parameters close to those of the resonance states reported in Ref.~\cite{2004NuPhA.745..155T}, which have undetermined total angular momentum $J$ and parity $\pi$. We assume that these twin resonance states correspond to the $4^+$ resonance states. A similar conclusion was drawn in Ref.~\cite{1994NuPhA.573...28D}. 

The results presented in Table~\ref{Tab:4PRHNPvsModel} show that, without cluster polarization, the $4^+$ resonance states have high energies (greater than 10 MeV) and large widths (about 3 MeV). 
Incorporating cluster polarization in both 6Ch and 8Ch approximations reduces the energies of these resonance states by a factor of two and significantly decreases their widths—by almost 100 times, from 2.9 MeV and 3.2 MeV to 34 keV and 8 keV, respectively.
Table~\ref{Tab:4PRHNPvsModel} also shows that the exotic channels $^{6}$Be+$^{2}n$ and $^{6}$He+$^{2}p$ play an insignificant role in the formation of the twin $4^+$ resonance states.

\begin{table}[tbp] \centering
\caption{Energies and widths of the twin 4$^+$ resonance states, obtained in different approximations. Experimental values from Ref.~\cite{2004NuPhA.745..155T} are given for comparison.} 
\begin{tabular}
[c]{|c|c|c|c|c|c|c|}\hline
& Model & Polarization &  $E$, MeV & $\Gamma$, keV & $E$,
MeV & $\Gamma$, keV\\\hline
4$^{+}$ & 6Chs & No &  10.330 & 2941.2 & 10.781 & 3203.5 \\
& 6Chs & Yes &  5.191 & 34.2 & 5.572 & 7.8\\\cline{2-7}
& 8Chs & No &  10.329 & 2941.2 & 10.781 & 3202.5\\
& 8Chs & Yes &  5.190 & 33.9 & 5.567 & 7.7\\\cline{2-7}
& Exp & - &  5.376 & 100.0 & 5.726 & 230.0\\\hline
\end{tabular}
\label{Tab:4PRHNPvsModel}
\end{table}
The spectra of the $1^-$ and $2^-$ resonance states obtained using different approximations are presented in Table~\ref{Tab:1M2MRHNPvsModel}. As shown in Table~\ref{Tab:1M2MRHNPvsModel}, our results closely match the experimental data for all the listed resonances. Interestingly, the positions of the $1^-$ and $2^-$ resonances are insensitive to the degree of polarization of the binary subsystems and to the inclusion of the $^{2}n+^6$Be and $^{2}p+^6$He channels, although the widths are influenced by these two factors.

The $2^-$ resonance and the lowest $1^-$ resonance can be associated with neutron scattering on the ground state and the first excited state of $^7$Be, respectively, while the second $1^-$ resonance can be interpreted as $^3$He scattering on $^5$He. This interpretation is supported by the fact that its resonance energy is very close to the corresponding decay thresholds of $^8$Be.

\begin{table}[tbp] \centering
\caption{Energies and widths of the 1$^-$  and 2$^-$ resonance states, obtained in different approximations. Experimental values from Ref.~\cite{2004NuPhA.745..155T} are shown for comparison.}
\begin{tabular}
[c]{|c|c|c|c|c|c|c|}\hline
& Model & Polarization &  $E$, MeV & $\Gamma$, keV & $E$,
MeV & $\Gamma$, keV\\\hline
1$^{-}$ & 6Chs & No &  2.141 & 1257.80 & 4.7163 & 8412.90\\
& 6Chs & Yes  & 2.141 & 299.61 & 4.724 & 4942.98\\\cline{2-7}
& 8Chs & No  & 2.141 & 35.11 & 4.4163 & 8254.39\\
& 8Chs & Yes  & 2.142 & 317.19 & 4.736 & 4783.28\\\cline{2-7}
& Exp & - &  2.146 & $\approx$650 & 4.746 & $\approx$4000\\\hline\hline
2$^{-}$ & 6Chs & No  & 1.694 & 360.00 &  & \\
& 6Chs & Yes  & 1.693 &  221.04  &  & \\\cline{2-7}
& 8Chs & No & 1.693 & 553.86  &  & \\
& 8Chs & Yes &  1.693 & 312.80 &  & \\\cline{2-7}
& Exp & - &  1.656 & 122 &  & \\\hline
\end{tabular}
\label{Tab:1M2MRHNPvsModel}
\end{table}

\subsection{Partial decay widths of the $4^+$, $1^+$, and $3^+$ twin states}

To gain deeper insight into the nature of the observed resonance states, we analyze the partial widths of each resonance, which allows us to identify the dominant decay channels. In Table~\ref{Tab:4PWidthsHNP}, we present the total and partial widths of the two $4^+$ resonance states, listing only the channels with partial widths above 1\%.

Table~\ref{Tab:4PWidthsHNP} shows that the first $4^+$ resonance most likely decays via the $\alpha+\alpha$ channel, while the second $4^+$ resonance has comparable decay probabilities through the $\alpha+\alpha$, $n+^7$Be, $^3$He+$^5$He, and $^3$H+$^5$Li channels. The partial width for the $^7$Li+$p$ channel is about 9\% for both $4^+$ resonances.

\begin{table}[tbph] \centering
\caption{Total and partial decay widths of the $4^+$ resonance states. Each row corresponds to a specific binary decay channel (BC), with $\Gamma_i$ (in keV) and $\Gamma_i/\Gamma$ (in \%) denoting the partial width and the branching ratio, respectively. $j_2$ and $j_1$ denote the angular momenta of the two-cluster subsystem and the relative motion of the third cluster, respectively.}
\begin{tabular}
[c]{|c|c|c|c|c|c|c|c|c|c|}\hline
\multicolumn{5}{|c|}{$E$=5.190 MeV, $\Gamma$=33.9 keV} &
\multicolumn{5}{|c|}{$E$=5.567 MeV, $\Gamma$=7.7 keV}\\\hline
 $\Gamma_{i}$ & $\Gamma_{i}/\Gamma$, \% & BC & $j_{1}$ & $j_{2}$ &
$\Gamma_{i}$ & $\Gamma_{i}/\Gamma$, \% & BC & $j_{1}$ & $j_{2}$\\\hline
 28.80 & 84.92 & $^{4}$He+$^{4}$He & 4 & 0 & 1.85 & 24.03 & $^{4}$He+$^{4}$He & 4 & 0\\
 2.94 & 8.67 & $p+^{7}$Li & 7/2 & 3/2 & 1.76 & 22.84 & $n+^{7}$Be & 7/2 & 3/2\\
 1.15 & 3.39 & $^{3}$H+$^{5}$Li & 7/2 & 3/2 & 1.30 & 16.87 & $^{3}$He+$^{5}$He & 7/2 & 3/2\\
 0.47 & 1.38 & $p+^{7}$Li & 7/2 & 1/2 & 1.06 & 13.70 & $^{3}$H+$^{5}$Li & 7/2 & 3/2\\
 0.38 & 1.11 & $n+^{7}$Be & 7/2 & 3/2 & 0.71 & 9.20 & $p+^{7}$Li & 7/2 & 3/2\\
&  &  &  &  & 0.38 & 4.88 & $n+^{7}$Be & 7/2 & 1/2\\
&  &  &  &  & 0.37 & 4.77 & $p+^{7}$Li & 7/2 & 1/2\\
&  &  &  &  & 0.10 & 1.25 & $^{3}$H+$^{5}$Li & 7/2 & 1/2\\\hline
\end{tabular}
\label{Tab:4PWidthsHNP}
\end{table}%
In Table~\ref{Tab:1PWidthsHNP}, we present the total and partial widths of the twin $1^+$ resonance states. Note that 22 channels—five partitions listed in Eq.~(\ref{eq:N002}), excluding the $^4$He+$^4$He channel, and all possible combinations of partial angular momenta $j_{1}$ and $j_{2}$—contribute to the formation of the $1^+$ resonance states. However, only three channels are open in the energy region near the first resonance state, while five channels are open at the energy of the second resonance state.

Two $p+^{7}$Li$(3/2^+)$ channels, with proton relative angular momenta $j_{1} = 1/2$ and $j_{1} = 3/2$, are the dominant decay modes for the lowest $1^+$ resonance state. At first glance, this result appears to contradict Fig.~\ref{Fig:Resons1PHNPvsChs}, where we demonstrated the importance of the $n+^{7}$Be and $^{3}$He+$^{5}$He channels in determining the position of the $1^+$ resonance state. These channels are indeed crucial; however, due to their higher threshold energies, they do not contribute to the decay of these resonance states.
\begin{table}[tbp] \centering
\caption{Total and partial decay widths of the $1^+$ resonance states. See the caption of Table~\ref{Tab:4PWidthsHNP} for definitions of the columns.}
\begin{tabular}
[c]{|c|c|c|c|c|c|c|c|c|c|}\hline
\multicolumn{5}{|c|}{$E$=0.385 MeV, $\Gamma$=9.63 keV} &
\multicolumn{5}{|c|}{$E$=1.237 MeV, $\Gamma$=350.90 keV}\\\hline
 $\Gamma_{i}$ & $\Gamma_{i}/\Gamma$, \% & BC & $j_{1}$ & $j_{2}$ & $\Gamma_{i}$ & $\Gamma_{i}/\Gamma$, \% & BC & $j_{1}$ & $j_{2}$\\\hline
 7.13 & 74.02 & $p$+$^{7}$Li & 1/2 & 3/2 & 230.51 & 65.69 & $p$+$^{7}$Li & 1/2 & 3/2\\
 2.50 & 25.98 & $p$+$^{7}$Li & 3/2 & 3/2 & 92.25 & 26.29 & $p$+$^{7}$Li & 3/2 & 3/2\\
 0.00 & 0.0 & $p$+$^{7}$Li & 5/2 & 3/2 & 14.38 & 4.10 & $p$+$^{7}$Li & 1/2 & 1/2\\
  &  &  &  &  & 13.77 & 3.92 & $p$+$^{7}$Li & 3/2 & 1/2\\
  &  &  &  & & 0.00 & 0.00 & $p$+$^{7}$Li & 5/2 & 3/2\\
\hline
\end{tabular}
\label{Tab:1PWidthsHNP}
\end{table}%
In Table \ref{Tab:3PWidthsHNP}, we display the total and partial widths of the
twin 3$^{+}$ resonance states. It can be seen that the first resonance state is
formed by a single binary channel, $p$+$^{7}$Li, with $^{7}$Li in the ground state, which accounts for 99.95\% of all decays. 
The second $3^+$ resonance state is primarily formed and decays through the $p+^{7}$Li and $n+^{7}$Be channels, both contributing almost equally.

\begin{table}[tbp] \centering
\caption{Total and partial decay widths of the 3$^+$ resonance states. See the caption of Table~\ref{Tab:4PWidthsHNP} for definitions of the columns.} 
\begin{tabular}
[c]{|c|c|c|c|c|c|c|c|c|c|}\hline
\multicolumn{5}{|c|}{$E$=1.671 MeV, $\Gamma$= 280.91 keV} &
\multicolumn{5}{|c|}{$E$=2.339 MeV, $\Gamma$= 260.56 keV}\\\hline
$\Gamma_{i}$ & $\Gamma_{i}/\Gamma$, \% & BC & $j_{1}$ & $j_{2}$ &
$\Gamma_{i}$ & $\Gamma_{i}/\Gamma$, \% & BC & $j_{1}$ & $j_{2}$\\\hline
280.77 & 99.95 & $p$+$^{7}$Li & 3/2 & 3/2 & 133.26 & 51.14 & $p$+$^{7}$Li & 3/2 & 3/2\\
0.10 & 0.04 & $p$+$^{7}$Li & 7/2 & 3/2 & 127.20 & 48.82 & $n$+$^{7}$Be & 3/2 & 3/2\\
0.03 & 0.01 & $p$+$^{7}$Li & 7/2 & 1/2 & 0.04 & 0.02 & $p$+$^{7}$Li & 7/2 & 3/2\\
0.01 & 0.00 & $p$+$^{7}$Li & 5/2 & 3/2 & 0.03 & 0.01 & $p$+$^{7}$Li & 5/2 & 3/2\\
 &  &  & &  & 0.02 & 0.01 & $p$+$^{7}$Li & 5/2 & 1/2\\\hline
\end{tabular}
\label{Tab:3PWidthsHNP}
\end{table}

\subsection{Formation of the twin $2^+$ Shape and Feshbach Resonances}
\label{sec:feshbach}

In this section, we analyze the formation mechanisms of shape and Feshbach resonances in a coupled two-channel system. We consider the $2^+$ resonance states located below and above the $p+^{7}$Li threshold. For simplicity, this analysis is carried out using a two-channel approximation (2C), where only the $p+^{7}$Li and $^{4}$He+$^{4}$He channels are taken into account. 
As shown in Fig.~\ref{Fig:Resons2PHNPvsChs} (see Section \ref{sec:resonance_parameters}), this approximation results in two resonance states—one below and one above the $p+^{7}$Li threshold.

To establish the nature of the obtained resonance states and evaluate the role of interchannel coupling in resonance formation, we introduce a scaling factor $f_{C}$, which controls the interchannel coupling in our model. In symbolic form, the modified equation can be written as:

\begin{equation}
\left(
\begin{array}
[c]{cc}%
\widehat{H}_{11} & f_{C}\widehat{H}_{12}\\
f_{C}\widehat{H}_{21} & \widehat{H}_{22}%
\end{array}
\right)  \left(
\begin{array}[c]{c}
\Psi_{1}\\
\Psi_{2}
\end{array}
\right)  =E\left(
\begin{array}[c]{cc}
N_{11} & f_{C}N_{12}\\
f_{C}N_{21} & N_{22}
\end{array}
\right)  \left(
\begin{array}[c]{c}
\Psi_{1}\\
\Psi_{2}
\end{array}
\right)  ,\label{eq:C001}
\end{equation}
where $\widehat{H}_{11}$ and $\widehat{H}_{22}$ describe the interactions within the $^{7}$Li+$p$ and $^{4}$He+$^{4}$He channels, respectively, while $\widehat{H}_{12}$ and $\widehat{H}_{21}$ govern the interchannel coupling. The off-diagonal Hamiltonians $\widehat{H}_{12}$ and $\widehat{H}_{21}$ define the dynamic coupling between $p+^{7}$Li and $^{4}$He+$^{4}$He, while the norm kernels $N_{12}$ and $N_{21}$ account for kinematic coupling effects. The diagonal norm kernels $N_{11}$ and $N_{22}$ include the influence of the Pauli principle in the corresponding channels.

For $f_C = 0$, both dynamic and kinematic couplings are switched off. In this uncoupled limit, the $p+^{7}$Li channel supports several bound states below its threshold and some resonance states above it, while the $^{4}$He+$^{4}$He channel produces a single shape resonance  near its threshold. 

By gradually increasing the coupling factor $f_C$ from 0 to 1, we can trace the evolution of the resonance states, starting from weak coupling and reaching the full interchannel coupling strength that corresponds to the actual interaction used in Fig.~\ref{Fig:Resons2PHNPvsChs} for the $2^+$ states. This approach allows us to identify which closed-channel bound states contribute to each observed resonance and to understand the role of configuration mixing during their formation.

\begin{figure}[ptb]
\begin{center}
\includegraphics[height=12.2572cm, width=13.518cm]{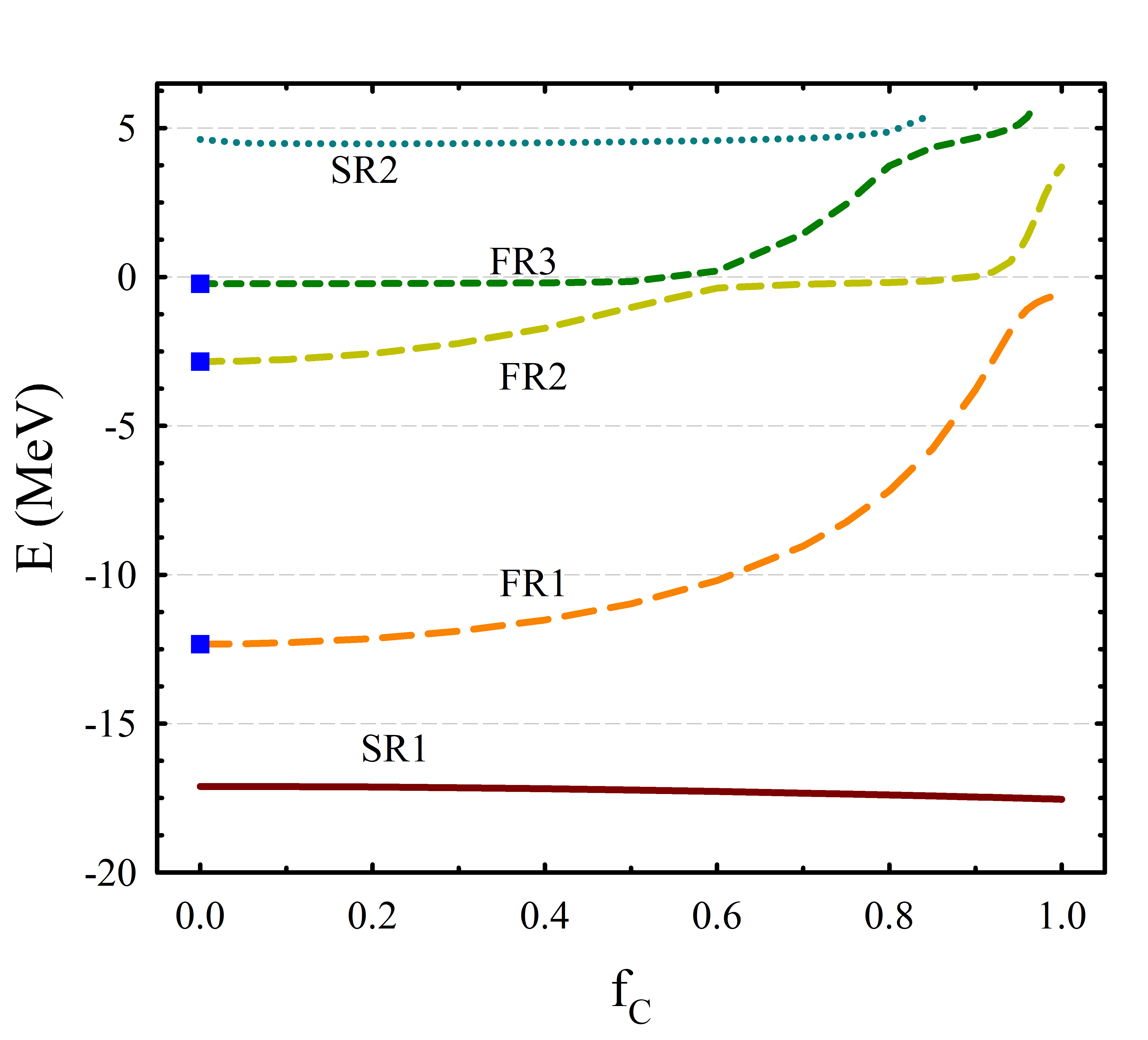}
\caption{Energies of the $2^+$ shape (SR1 and SR2) and Feshbach (FR1, FR2, and FR3) resonance states as a function of the coupling parameter $f_C$. Blue boxes indicate the positions of bound states in the $^7$Li + $p$ channel at $f_C = 0$.}
\label{Fig:FeshbResons2PvsCCHNP}
\end{center}
\end{figure}

In Fig.~\ref{Fig:FeshbResons2PvsCCHNP}, we demonstrate how the $2^+$ resonance states are formed. If the coupling constant $f_C = 0$, the $^{4}$He+$^{4}$He channel generates one well-known $2^+$ resonance state (SR1), located near the $^{4}$He+$^{4}$He threshold.  Additionally, one shape resonance state (SR2) and three bound states (which later evolve into Feshbach resonances, FR1, FR2, and FR3) are formed in the $^{7}$Li+$p$ channel when $f_C = 0$. As $f_C$ increases, these bound states evolve into resonances as coupling to the open $^{4}$He+$^{4}$He channel allows them to decay.

By slightly increasing $f_C$, we obtain five resonance states, meaning that the bound states in the $^{7}$Li+$p$ channel are transformed into narrow resonances. Consequently, these resonances can be considered Feshbach resonances, which occur when a state primarily associated with a closed-channel configuration interacts with an open-channel configuration, forming a resonance in the continuum.

The rate of energy shift provides direct insight into the strength of interaction between the channels. A smooth shift in resonance energy with increasing coupling strength indicates a weak to moderate coupling regime, where the resonance remains primarily influenced by the closed-channel bound state.  
As the coupling becomes stronger, the resonance energy shifts significantly, and the state acquires characteristics similar to a shape resonance.

The parameter $f_{C}$ defines three coupling regions: weak, intermediate, and strong. As shown in Fig.~\ref{Fig:FeshbResons2PvsCCHNP}, these regions differ for the first, second, and third Feshbach resonance states (FR1, FR2, and FR3). The third Feshbach resonance state (FR3), located near the $^{7}$Li+$p$ threshold, exhibits the largest weak-coupling region, extending from $f_{C} \approx 0$ to $f_{C} \approx 0.5$. 
Fig.~\ref{Fig:FeshbResons2PvsCCHNP} also indicates that for $f_{C} > 0.6$, the system enters the strong-coupling regime, where increasing $f_{C}$ leads to substantial changes in the parameters of the first (FR1) and third (FR3) Feshbach resonance states.

Notably, as the coupling strength increases, the role of the near-threshold resonance is successively transferred from FR3 to FR2 and then to FR1, as evidenced by the avoided crossings in Fig.~\ref{Fig:FeshbResons2PvsCCHNP}. This behavior reflects a resonance identity switching mechanism, in which different closed-channel states dominate the near-threshold resonance at different values of the coupling parameter. Despite these internal structural changes, the energy of the near-threshold resonance remains nearly constant across a wide range of $f_C$, indicating strong threshold pinning due to its coupling to the open $^4$He+$^4$He channel. While avoided crossings are typically associated with strong coupling, the minimal energy shift and narrow width of the near-threshold resonance suggest that the actual coupling to the open $^4$He+$^4$He channel remains relatively weak. Thus, the avoided crossings in this case primarily reflect competition among closed-channel states to couple to the continuum, rather than strong mixing with the open channel itself.

The shape resonance state SR2, located above the $^{7}$Li+$p$ threshold, undergoes only slight parameter changes. As seen in Fig.~\ref{Fig:FeshbResons2PvsCCHNP}, its position remains stable in the region $0 < f_{C} < 0.75$, which suggests that its properties are primarily determined by the $^{7}$Li+$p$ channel rather than by coupling to $^{4}$He+$^{4}$He. This is expected because shape resonances are primarily determined by the centrifugal and potential barriers in the open channel, making them less sensitive to interchannel coupling.

\begin{figure}[ptbh]
\begin{center}
\includegraphics[height=16.1825cm,width=13.5707cm]{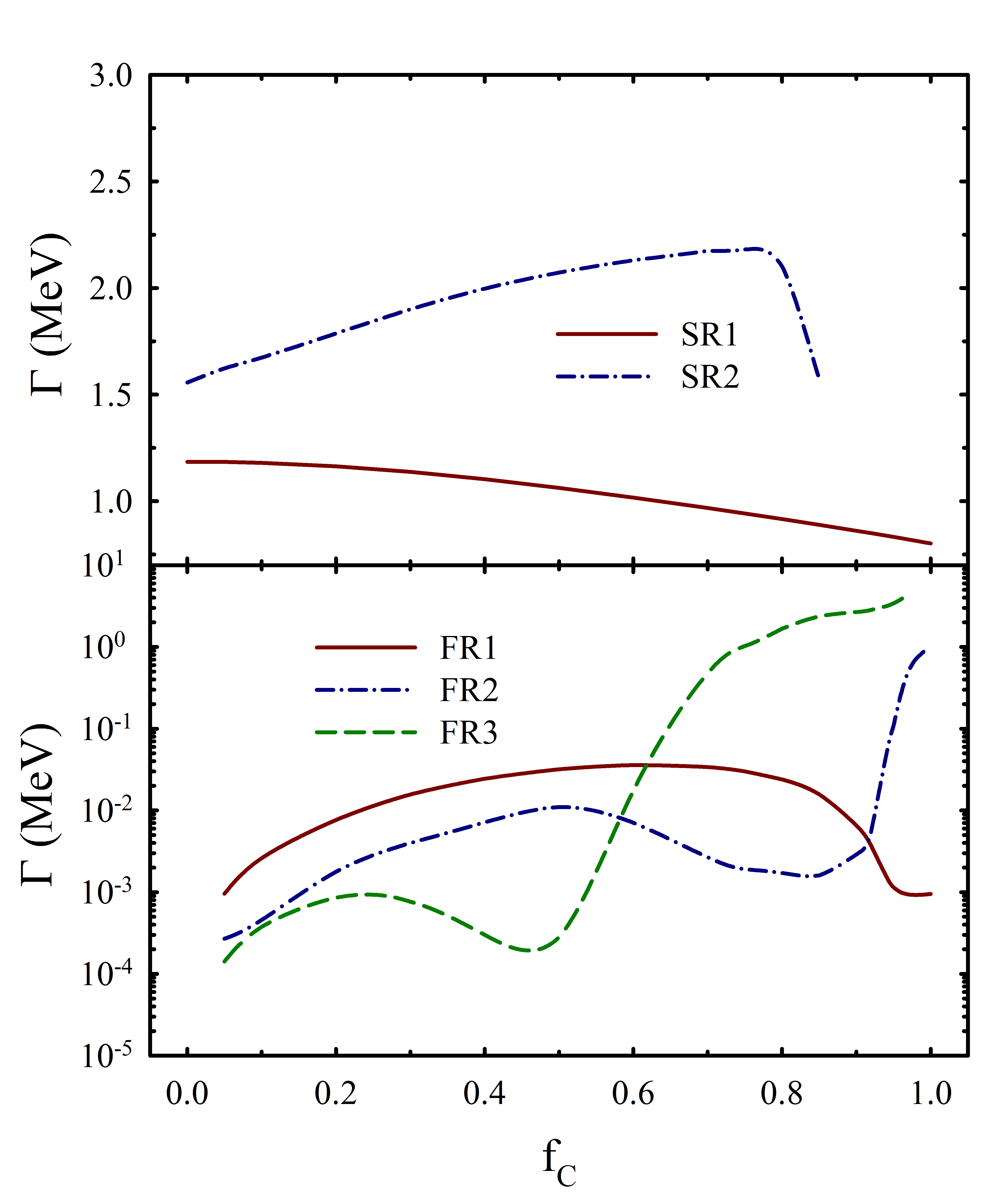}
\caption{Widths of the $2^+$ shape (SR1, SR2) and Feshbach (FR1, FR2, FR3) resonance states as a function of the coupling constant $f_C$.}\label{Fig:FeshbResons2PWvsCCHNP}
\end{center}
\end{figure}

Fig.~\ref{Fig:FeshbResons2PWvsCCHNP} demonstrates the dependence of the width of shape 
(the upper panel) and Feshbach (the lower panel) resonance states on the coupling constant $f_C$. 
With increasing $f_C$, the width of the first shape resonance (SR1) gradually decreases.
This suggests that interchannel coupling modifies the open-channel potential in a way that reduces its decay probability, effectively stabilizing the resonance. 
The width of the second shape resonance (SR2) generally increases with $f_C$ until reaching a critical point at 
$f_C \approx 0.75$, where it undergoes a sharp drop, possibly indicating a transition in its resonance nature.

The widths of all Feshbach resonance states start from $\Gamma = 0$ at $f_C = 0$, 
as they originate as bound states. With increasing $f_C$, they reach maxima, marking the 
transition from bound states to resonances. Beyond these maxima, the FR1 resonance experiences 
width reduction, while the FR2 and FR3 Feshbach resonances transform into shape resonances.

\begin{figure}[ptbh]
\begin{center}
\includegraphics[height=12.0111cm,width=13.406cm]{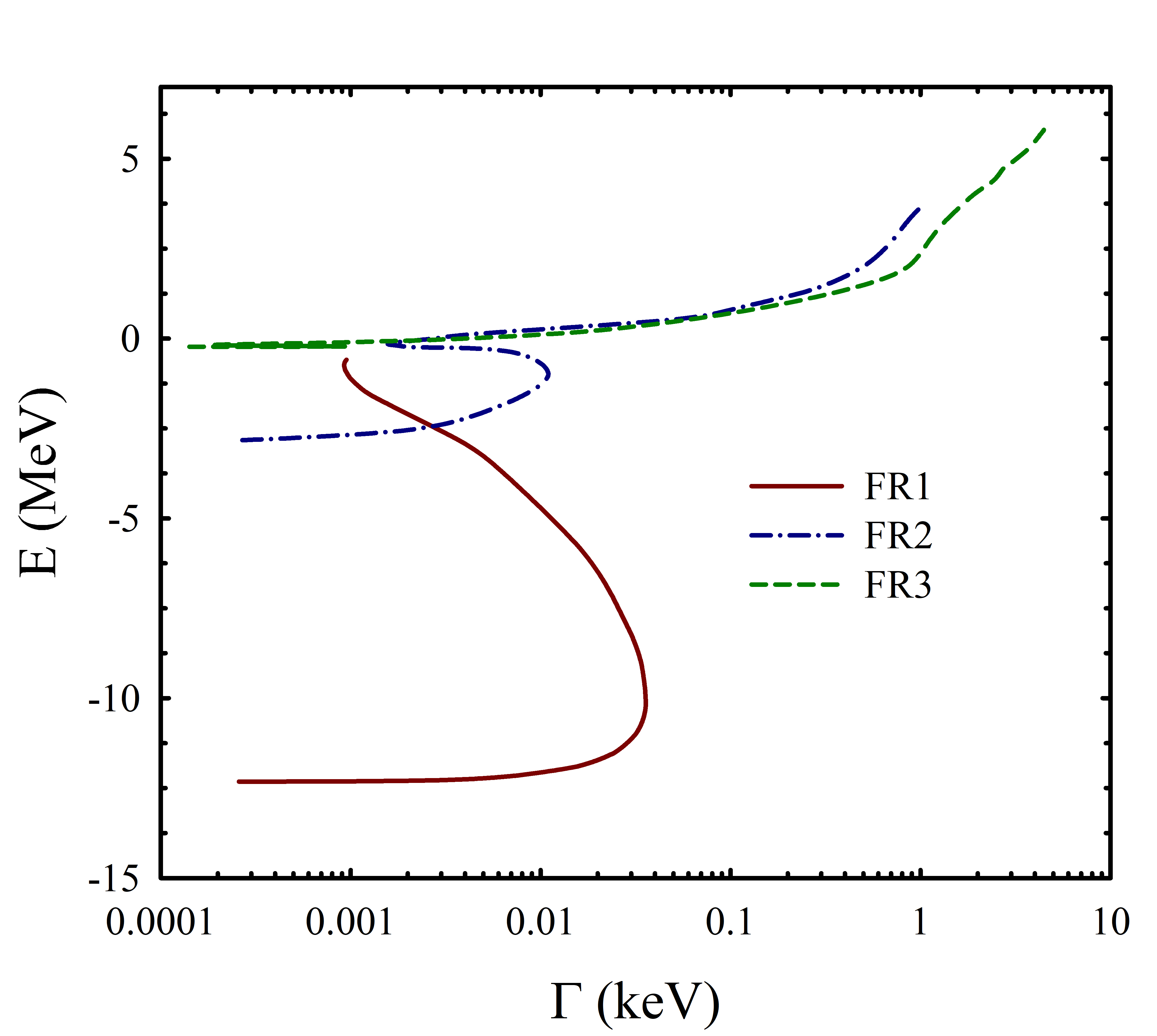}
\caption{Trajectories of the Feshbach $2^+$ resonance states in the energy-width plane as a function of the coupling constant $f_C$.}
\label{Fig:EvsG2PFRes}
\end{center}
\end{figure}

As observed in Fig.~\ref{Fig:FeshbResons2PvsCCHNP}, the FR2 and FR3 Feshbach resonance 
states undergo a transition to shape resonances at critical points $f_C = 0.9$ and 
$f_C = 0.6$, respectively. At these points, their widths increase sharply, reflecting a 
shift from a closed-channel-dominated behavior to an open-channel-dominated resonance.
Another perspective on the trajectories of Feshbach $2^+$ resonance states is presented in Fig.~\ref{Fig:EvsG2PFRes}, where the trajectories are displayed in the energy-width plane.  
All trajectories are orthogonal to the energy axis at their starting points, where the coupling constant $f_{C}$ is small. In this region, the energies of the resonance states decrease slowly, while their widths increase rapidly. Fig.~\ref{Fig:EvsG2PFRes} explicitly demonstrates the oscillatory behavior of the widths of the first (FR1) and second (FR2) resonance states.

It is worthwhile to note that the behavior of the trajectory of the $2^+$ resonance states aligns with the general properties of Feshbach resonance states. For example, a similar dependence of energy and width on the coupling constant was obtained in Ref.~\cite{1997JPhA...30.5543V}, where a simple two-channel model was used to study resonance states arising from the coupling of two channels with different threshold energies. In this model, square-well potentials simulated the interaction within each channel as well as the coupling between them. Varying the coupling constant from zero to large values produced similar trajectories for the Feshbach resonance states.

This analysis reveals that the near-threshold resonance observed at \( f_C = 1 \) originates from a bound state in the $^7$Li + $p$ channel, and acquires its resonant nature through coupling to the $^4$He + $^4$He continuum. Its formation involves configuration mixing via avoided crossings, yet its narrow width and weak energy shift confirm its predominantly Feshbach character.

\subsection{Phase shifts and eigenphase shifts for $p$+$^{7}$Li and $n$+$^{7}$Be scattering in the $1^+$, $1^-$, and $2^-$ states}

In this section, we consider how resonance states manifest themselves in scattering parameters. To this end, we use two related representations.
In the first representation, we study the behavior of the diagonal matrix elements of the scattering $S$-matrix, given by
\[
S_{cc}=\eta_{cc}\exp\left\{  2i\delta_{cc}\right\}  ,
\]
where $\delta_{cc}$ is the phase shift for elastic scattering in channel $c$, and $\eta_{cc}$ is the inelasticity parameter. By reducing the full $S$-matrix $\left\Vert S_{c\widetilde{c}} \right\Vert$ to diagonal form, we obtain the second representation, which consists of the eigenphase shifts $\delta_{\alpha}$.

The eigenphase shifts provide a complementary view of the resonance structure. Each resonance typically manifests as a rapid increase in a single eigenphase shift by approximately 180$^\circ$, indicating the dominant contribution of a specific scattering configuration. This representation is especially useful in the presence of coupled channels, as it isolates individual resonances that may overlap in the physical phase shifts.

\begin{figure}[ptb]
\begin{center}
\includegraphics[height=16.7053cm,width=13.5487cm]{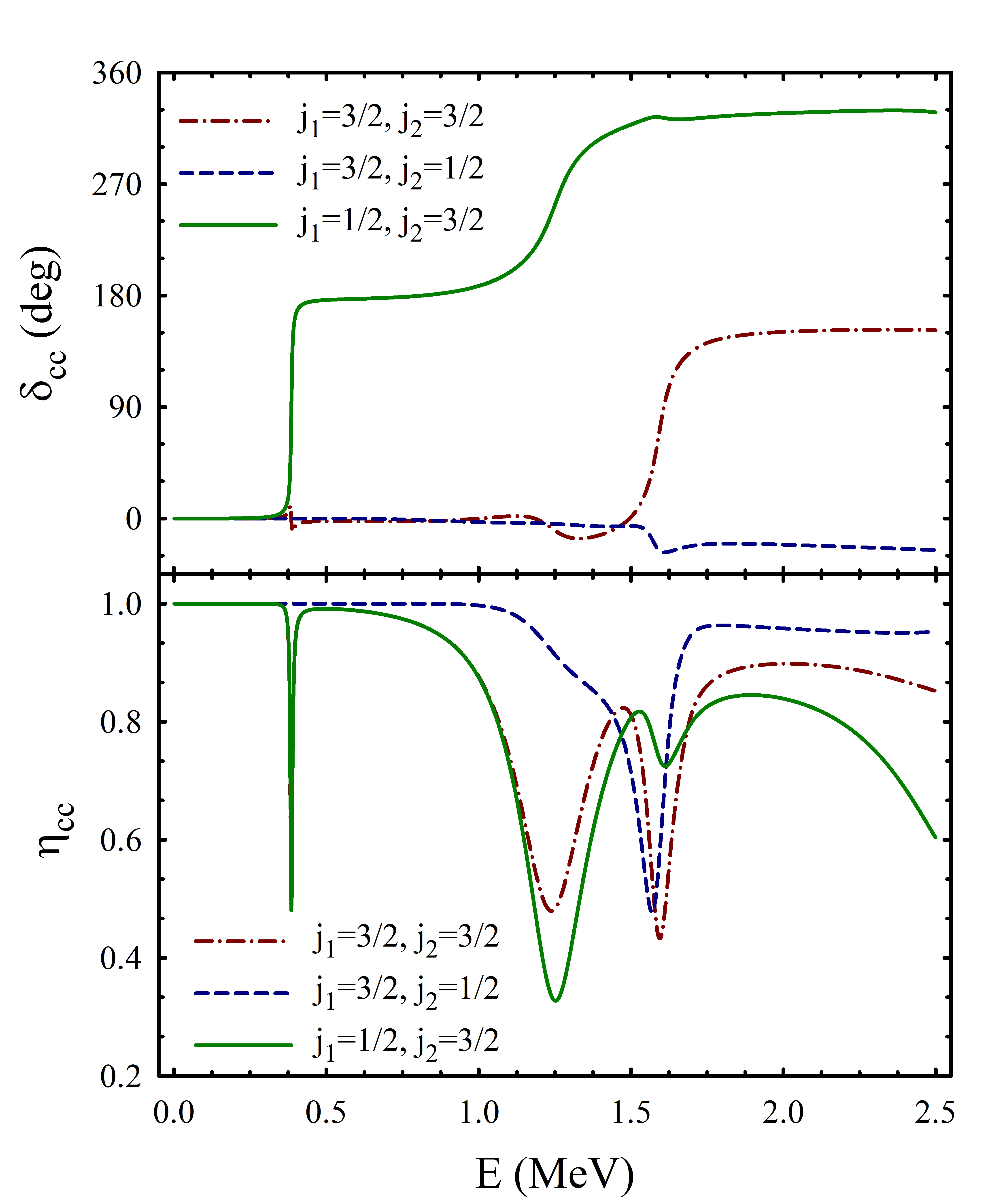}
\caption{Phase shifts $\delta_{cc}$ and inelasticity parameters $\eta_{cc}$ for $p$+$^{7}$Li scattering in the 1$^{+}$ state. $j_1$ denotes the angular momentum of the proton's relative motion, and $j_2$ denotes the angular momentum of $^{7}$Li, respectively.}
\label{Fig:PhasesEtas1P}
\end{center}
\end{figure}

In Fig.~\ref{Fig:PhasesEtas1P}, we show the phase shifts and inelasticity parameters for $p+^{7}$Li scattering in the $1^+$ state. The channels shown correspond to scattering of a proton off $^{7}$Li in its ground state ($j_{2} = 3/2$) and in its first excited state ($j_{2} = 1/2$). The phase shift describing scattering of a proton with angular momentum $j_{1} = 1/2$ on the ground state of $^{7}$Li exhibits resonance behavior at energies $E \approx 0.385$ MeV and $E \approx 1.237$ MeV. Evidently, the first resonance is very narrow, while the second is broader. The inelasticity parameter for the channel with quantum numbers $j_{1} = 1/2$ and $j_{2} = 3/2$ also exhibits a pronounced minimum at the resonance energies.

\begin{figure}[ptb]
\begin{center}
\includegraphics[height=16.4417cm,width=13.4543cm]{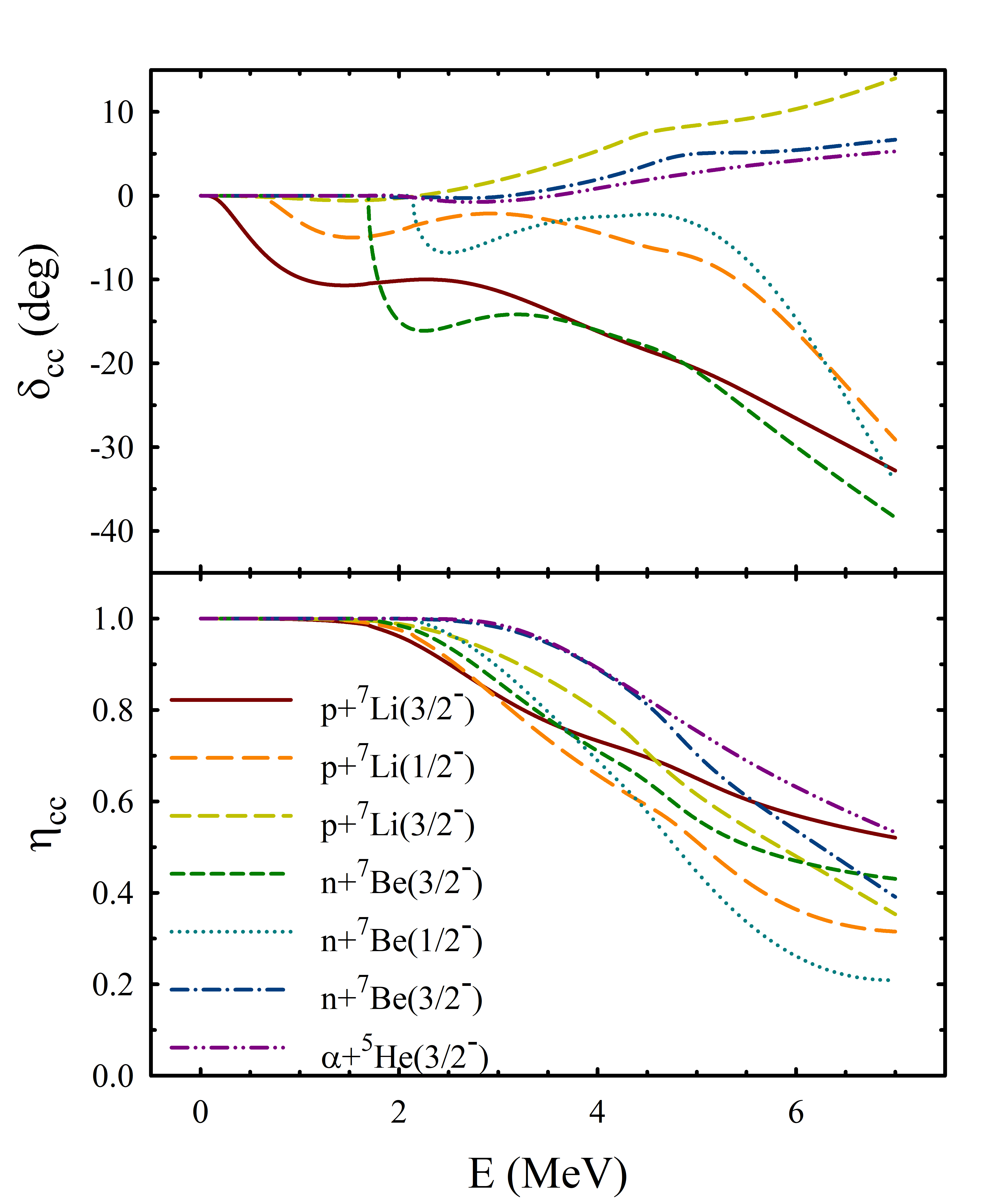}
\caption{Phase shifts $\delta_{cc}$ and inelasticity parameters $\eta_{cc}$ for $p+^7$Li, $n+^7$Be, and $\alpha+^5$He scattering in the $1^-$ state as functions of the energy $E$ above the $p+^7$Li threshold.}
\label{Fig:PhasesEtas1M}
\end{center}
\end{figure}

\begin{figure}[ptb]
\begin{center}
\includegraphics[height=12.1385cm,width=13.5728cm]{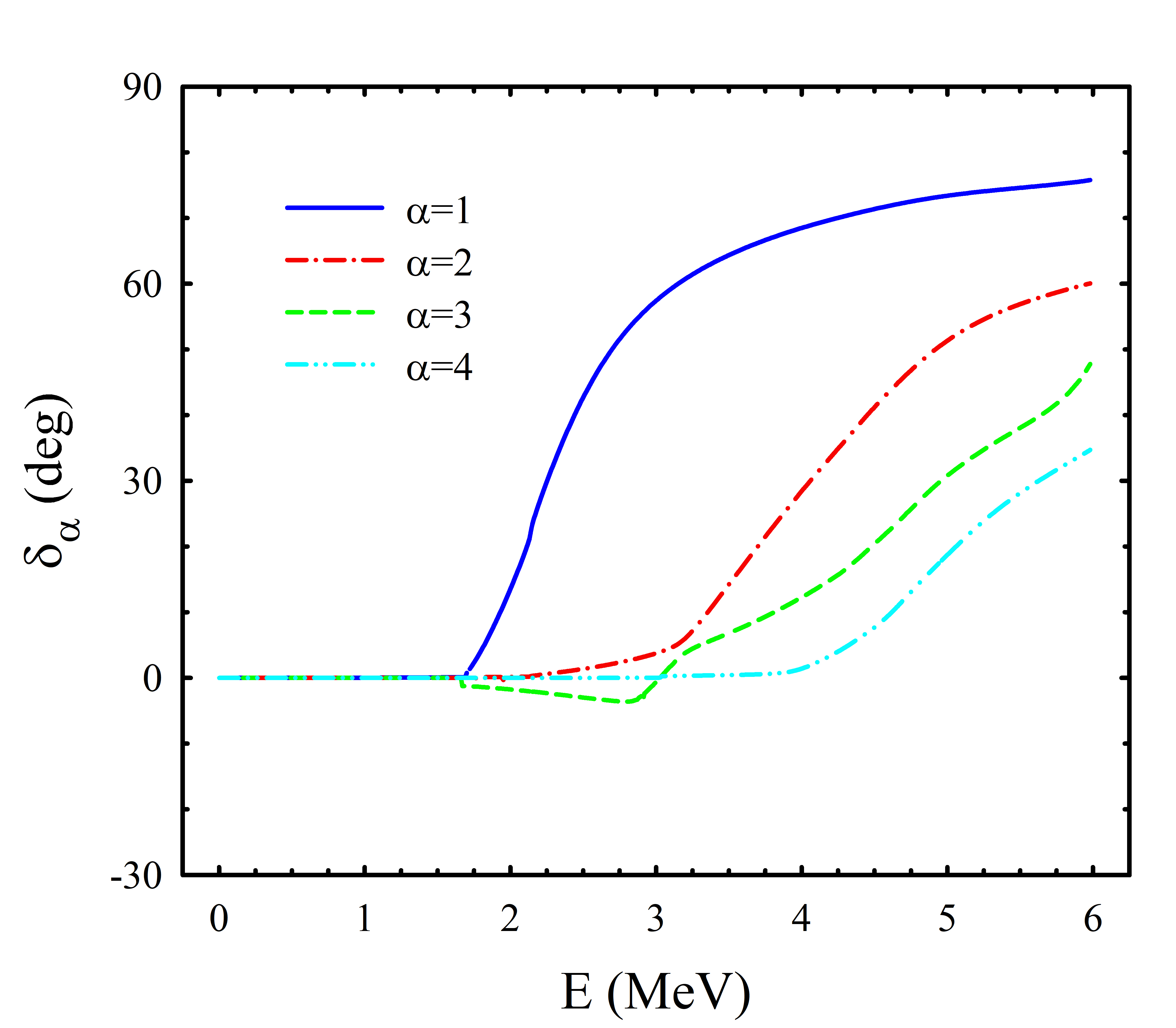}
\caption{Eigenphase shifts $\delta_{\alpha}$ for scattering in the 1$^{-}$ state.}
\label{Fig:EigenPhases1M}
\end{center}
\end{figure}

The scattering parameters for the $1^-$ and $2^-$ states exhibit quite different behavior. In Fig.~\ref{Fig:PhasesEtas1M}, we display the phase shifts and inelasticity parameters for the $1^-$ state. As pointed out above, we have identified two broad resonance states at energies $E = 2.142$~MeV and $E = 4.736$~MeV. However, we do not observe a rapid increase in the phase shifts or minima in the inelasticity parameters at these energies. There is only a slow rise of the phase shifts at energies above 3 MeV. To reveal the resonance structure in this case, we need to examine the eigenphase shifts. Indeed, some of the eigenphase shifts, shown in Fig.~\ref{Fig:EigenPhases1M}, exhibit resonance behavior characteristic of broad resonance states. Notably, the first, narrower resonance appears predominantly in one eigenphase shift, while the second, broader resonance is associated with a different eigenphase. This observation is consistent with the general feature of eigenphase shifts, where each resonance typically dominates one eigenchannel.

Near the neutron threshold, the shape and magnitude of the $^7$Li(p,p')$^7$Li* and $^7$Be(n,p)$^7$Li cross sections are closely tied to the behavior of the $2^-$ phase shift. Experimentally~\cite{brown1973polarization}, this phase shift exhibits a discontinuity in its slope near the threshold—known as the Wigner-Baz' cusp~\cite{1948PhRv...73.1002W, baz1958energy, baz1959resonance, 1959AdPhy...8..349B}—a hallmark of resonances with low-energy $s$-wave contributions in the neutron channel.

\begin{figure}[ptb]
\begin{center}
\includegraphics[height=3.9608in,width=4.4763in]{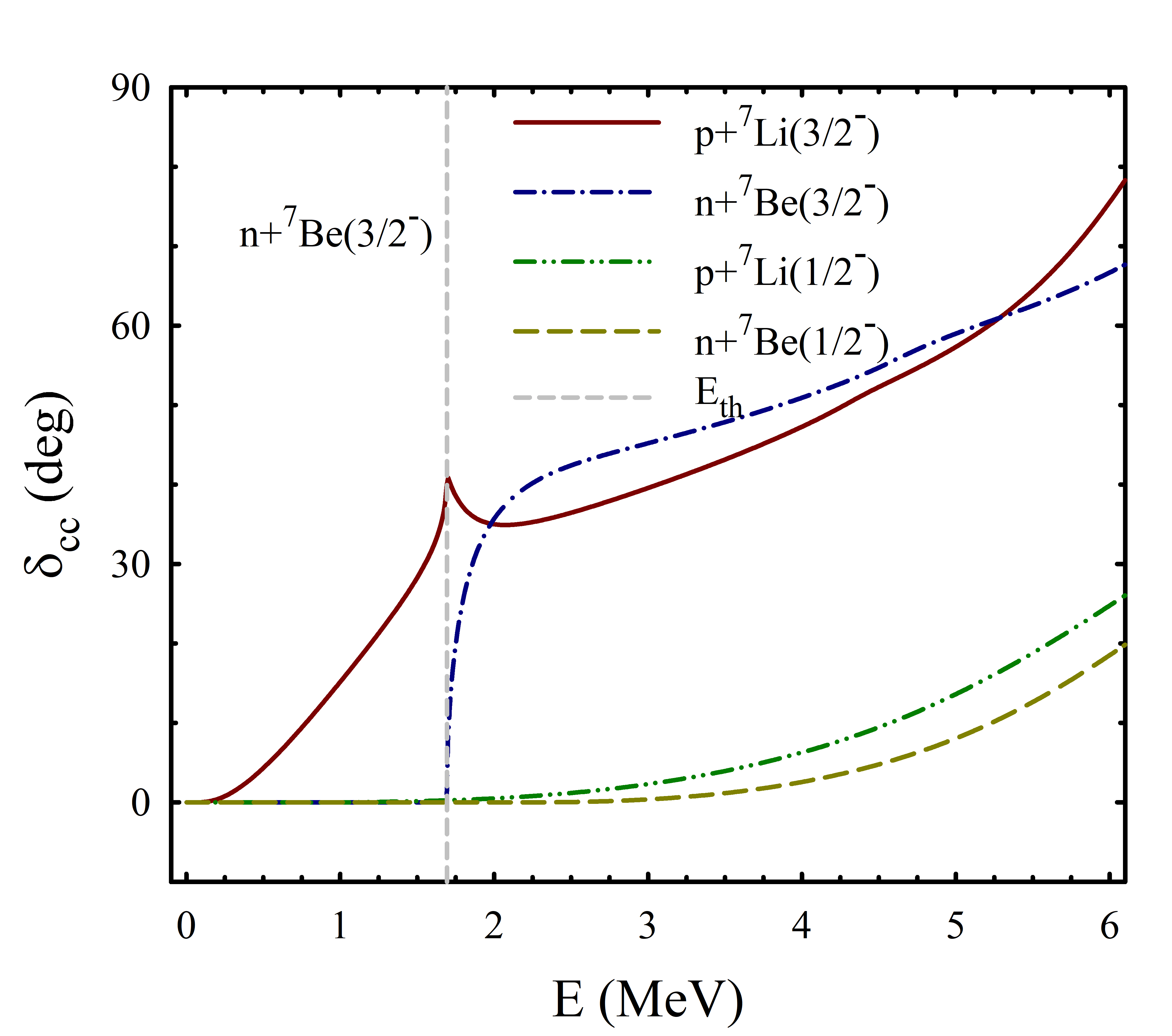}
\caption{Phase shifts $\delta_{cc}$ of the elastic $p$+$^{7}$Li and $n$+$^{7}$Be scattering in the 2$^{-}$ state. Vertical line indicates the threshold energy of the $n$+$^7$Be  channel}
\label{Fig:Phases2M}
\end{center}
\end{figure}

Several prominent $2^-$ phase shifts are shown in Fig.~\ref{Fig:Phases2M}, with the largest shifts corresponding to proton scattering on the ground state of $^7$Li and neutron scattering on the ground state of $^7$Be. Notably, the phase shift for $p+^7$Li(3/2$^-$) scattering clearly demonstrates the Wigner-Baz' cusp. However, the GCM calculation does not reproduce this experimental energy dependence~\cite{1994NuPhA.573...28D}, since the $2^-$ state obtained in the GCM appears as an almost pure proton resonance.

In Table~\ref{Tab:2MWidthsHNP}, we present the total and partial widths of the $2^-$ resonance state determined at the energy $E=1.693$~MeV. This resonance state decays predominantly into the $p+^{7}$Li channel, with $^7$Li in the ground state and proton angular momentum $j_{1} = 1/2$. Although the partial width into the $n+^7$Be channel is small, its coupling is crucial for the formation of the $2^-$ resonance, as discussed above. The opening of the $n+^7$Be channel involves an $s$-wave neutron, giving rise to the characteristic Wigner-Baz' cusp observed in the experimental phase shift.

\begin{table}[tbph] \centering
\caption{Total and partial decay widths of the 2$^-$ resonance state. See the caption of Table~\ref{Tab:4PWidthsHNP} for definitions of the columns.}
\begin{tabular}
[c]{|c|c|c|c|c|}\hline
\multicolumn{5}{|c|}{$E$= 1.693 MeV, $\Gamma$= 312.80 keV}\\\hline
$\Gamma_{i}$, keV & $\Gamma_{i}/\Gamma$, \% & BC & $j_{1}$ & $j_{2}%
$\\\hline
294.43 & 94.12 & $p+^{7}$Li & 1/2 & 3/2\\\hline
10.53 & 3.37 & $p+^{7}$Li & 5/2 & 3/2\\\hline
5.268 & 1.68 & $n+^{7}$Be & 1/2 & 3/2\\\hline
1.73 & 0.55 & $p+^{7}$Li & 5/2 & 1/2\\\hline
0.82 & 0.26 & $p+^{7}$Li & 3/2 & 3/2\\\hline
\end{tabular}
\label{Tab:2MWidthsHNP}
\end{table}

\subsection{Comparison with other models}

In Table~\ref{Tab:ComparMethods}, we summarize results for the resonance parameters of $^{8}$Be obtained using different microscopic methods and compare them with our calculations. We selected those models that provide both the energies and widths of resonance states. In Ref.~\cite{1994NuPhA.573...28D}, the three-cluster Generator Coordinate Method (GCM) was used to calculate energies and reduced widths, while in Ref.~\cite{Fernandez2024}, the Gamow Shell Model (GSM) provided both energies and total widths.

In Table~\ref{Tab:ComparMethods}, the experimental order of resonance states is used for consistency in comparison. A substantial portion of our results, particularly the resonance energies, show good agreement with the GSM calculations. According to experimental data, both the GSM and our model place the two $1^+$ resonance states above the $p+^7$Li threshold, whereas in the GCM, one of these states appears as a bound state below the threshold. Notably, the energies of the lowest $2^+$, $1^+$, and $3^+$ resonances obtained from the GSM and our model are very similar. However, the GSM overestimates the splitting of the $1^+$ doublet and underestimates the splitting of the $2^+$ doublet. Our model yields splittings of the $1^+$ and $2^+$ states that are closer to the experimental values, while it overestimates the splitting of the $3^+$ doublet.

Additionally, the GSM predicts a $4^+$ resonance approximately 2.5~MeV above the $^7$Li+$p$ threshold, which agrees well with the experimental energy, but this resonance is absent in both our model and the GCM. At the same time, Ref.~\cite{Fernandez2024} does not report the twin $4^+$ resonances slightly above 5~MeV, which are predicted by both our model and the GCM~\cite{1994NuPhA.573...28D}, and observed experimentally in Ref.~\cite{2004NuPhA.745..155T} as resonances without assigned spin and parity.
 In Ref.~\cite{1994NuPhA.573...28D}, the lower $4^+$ twin state is identified as a $T=1$ state, while the higher is assigned as $T=0$. 
 
 There is also a notable correspondence between the second $1^+$ and $3^+$ twin resonances in our model and those obtained with the GCM. The small values of the reduced alpha widths $\theta_\alpha^2$ for the lower $2^+$ and $4^+$ twin states obtained within the GCM are consistent with the narrow total widths of these states found in our calculations.

Overall, the reduced widths of the twin $2^+$, $1^+$, and $3^+$ resonances obtained within the GCM agree well with the partial widths of these states determined in our calculations (see Tables \ref{Tab:4PWidthsHNP}, \ref{Tab:1PWidthsHNP}, \ref{Tab:3PWidthsHNP}). In particular, the relatively large values of $\theta_p^2$ for the lower $2^+$, $3^+$, and also the $1^+$ state further support our conclusion about the dominant role of the $^7$Li$+p$ channel in the formation and decay of these resonances. At the same time, comparable values of $\theta_p^2$ and $\theta_n^2$ for the second $3^+$ state indicate nearly equal contributions from the $^7$Li$+p$ and $^7$Be$+n$ channels, consistent with our findings. Concerning the twin $4^+$ states, our results suggest a dominance of the $\alpha+\alpha$ channel in the lower state, while the second state decays into multiple channels. On the other hand, the GSM reduced widths suggest an alternative scenario, emphasizing instead the importance of the $\alpha+\alpha$ channel in the higher-energy state.

The $2^-$ resonance at 18.91~MeV plays a critical role in the $^7$Be(n,p)$^7$Li reaction, though discrepancies exist in the reported widths due to varying definitions of resonance parameters. While the compilation in Ref.~\cite{2004NuPhA.745..155T} lists a width of $\Gamma = 122$~keV, Descouvemont and Baye~\cite{1994NuPhA.573...28D} argue that larger values (e.g., $\Gamma \simeq 2$~MeV), consistent with a Breit-Wigner parametrization of cross sections, are also justified. In our model, the width of this resonance slightly exceeds 300~keV.

We also include in Table~\ref{Tab:ComparMethods} the results of Ref.~\cite{2024PhRvC.110a5503G}, obtained using the no-core shell model with continuum (NCSMC). This approach treats bound and unbound states within a unified framework, employing chiral two- and three-nucleon interactions proposed in Ref.~\cite{2003PhRvC..68d1001E}. The authors of Ref.~\cite{2024PhRvC.110a5503G} predict and analyze several low-energy resonances (e.g., $0^+$, $2^+$, and $4^+$) as well as high-energy $2^+$, $1^+$, and $3^+$ states. In Table~\ref{Tab:ComparMethods}, we include only the latter set for comparison. As shown in the table, the NCSMC calculations tend to overestimate the resonance energies of the twin states.
\begin{table}[htbp] \centering
\caption{Spectrum of resonance states in $^8$Be near the $p+^7$Li threshold, determined by different theoretical methods. Energies $E$ are given in MeV, and total widths $\Gamma$ are in keV. Dimensionless reduced widths $\theta_{\alpha}^{2}$, $\theta_{p}^{2}$, and $\theta_{n}^{2}$ for the $\alpha+\alpha$, $^7$Li$+p$, and $^7$Be$+n$ channels are given in percent, assuming a channel radius of 8.4~fm.}. \begin{tabular}
[c]{|c|c|c|c|c|c|c|c|c|c|c|}\hline
Method & \multicolumn{2}{|c|}{Present model} & \multicolumn{4}{|c}{GCM, \cite{1994NuPhA.573...28D}} & \multicolumn{2}{|c|}{GSM,
\cite{Fernandez2024}} & \multicolumn{2}{|c|}{NCSMC, \cite{2024PhRvC.110a5503G}}\\\hline
$J^{\pi}$ & $E$ & $\Gamma$ & $E$ & $\theta_{\alpha}^{2}$ & $\theta_{p}^{2}$ & $\theta_{n}^{2}$ & $E$ & $\Gamma$ & $E$ & $\Gamma$\\\hline
2$^{+}$ & -0.816 & 16.82 & -0.61 & 0.3 & 4.8 & 0.3 & -0.854 & 90 & -0.10 & -\\\hline
2$^{+}$ & -0.238 & 243.08 & -0.15 & 1.1 & 0.8 & 3.4 & -0.604 & 90 & 0.31 & -\\\hline
$p+^{7}$Li & 0.0 &  & 0.0 &  &  &  & 0.0 &  &  & \\\hline
1$^{+}$ & 0.385 & 9.63 & -0.07 &  & 2.3 & 1.1 & 0.336 & 9.1 & 0.734 & 89.9\\\hline
1$^{+}$ & 1.237 & 350.90 & 1.01 &  & 15.4 & 0.6 & 1.526 & 44 & 1.098 & 133.2\\\hline
2$^{-}$ & 1.693 & 0.313 &  &  &  &  &  &  &  & \\\hline
3$^{+}$ & 1.671 & 280.91 & 2.18 &  & 8.9 & 3.5 & 1.526 & 212 & 2.646 & 794.1\\\hline
3$^{+}$ & 2.339 & 260.56 & 2.89 &  & 7.2 & 9.8 & 2.026 & 204 & 2.868 & 389.4\\\hline
1$^{-}$ & 2.142 & 697.15 &  &  &  &  &  &  &  & \\\hline
4$^{+}$ & - & - &  &  &  &  & 2.496 & 987 &  & \\\hline
2$^{+}$ & 2.963 & 0.517 &  &  &  &  & 2.756 & 163 &  & \\\hline
1$^{-}$ & 4.736 & 4783.28 &  &  &  &  &  &  &  & \\\hline
4$^{+}$ & 5.190 & 34.0 & 5.49 & 0.6 & 0.2 & 0.1 &  &  &  & \\\hline
4$^{+}$ & 5.567 & 7.7 & 5.99 & 6.8 & 0.3 & 0.3 &  &  &  & \\\hline
1$^{-}$ & 6.768 & 6202 &  &  &  &  &  &  & & \\\hline
\end{tabular}
\label{Tab:ComparMethods}
\end{table}

\begin{figure}[ptbh]
\begin{center}
\includegraphics[height=12.1913cm,width=13.4829cm]{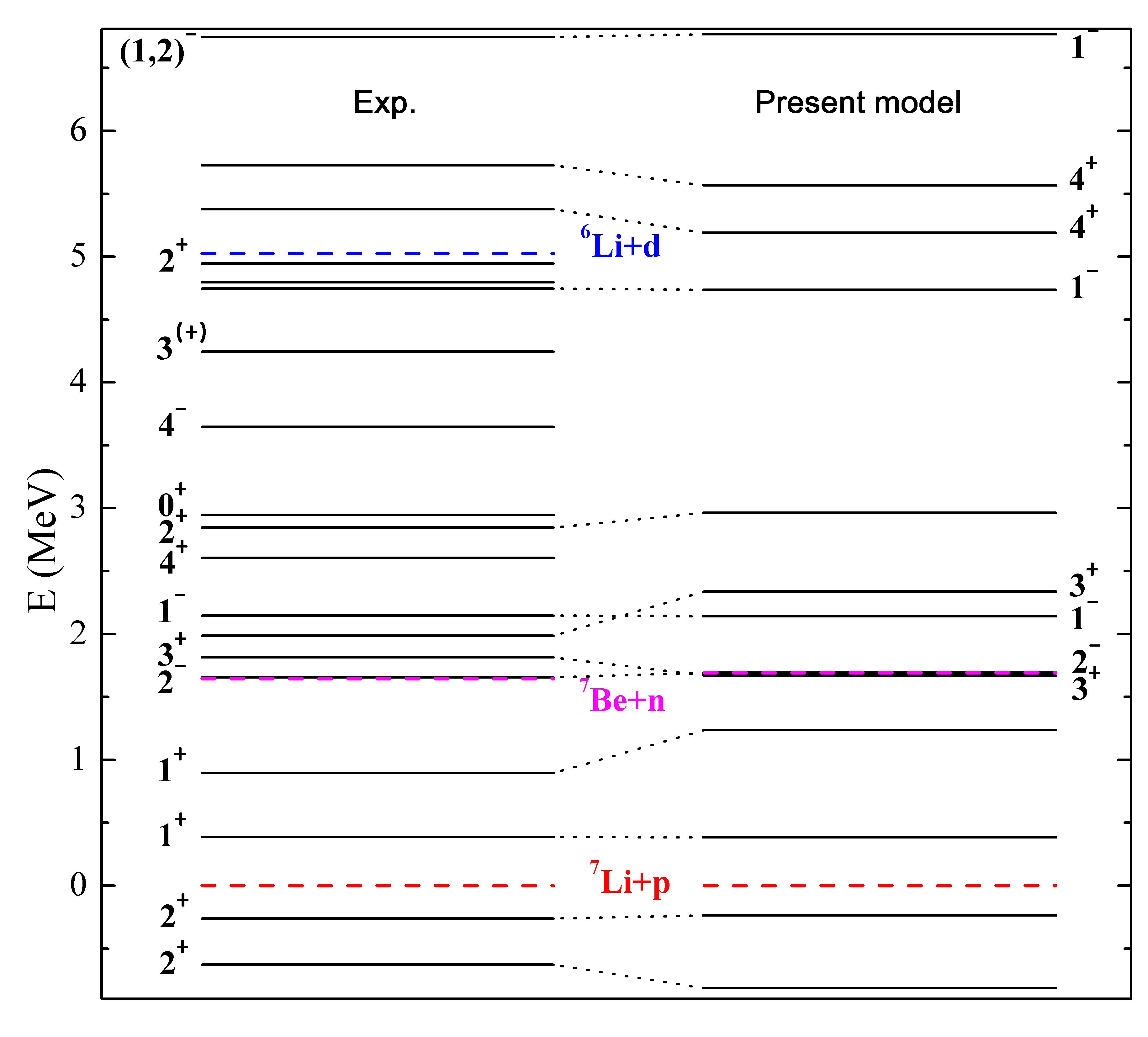}
\caption{Comparison of calculated spectra with experimental data from Ref.~\cite{2004NuPhA.745..155T} for high-energy resonance states in $^8$Be near the $p+^7$Li threshold. Dashed lines indicate the main binary decay thresholds.}
\label{Fig:SpectrRS8BeTvsE}
\end{center}
\end{figure}

In Fig.~\ref{Fig:SpectrRS8BeTvsE}, we summarize the spectrum of high-energy resonance states in $^{8}$Be obtained in our model and compare it with experimental data \cite{2004NuPhA.745..155T}. The figure demonstrates fairly good agreement between our results and the experimental values.

\section{Conclusions \label{Sec:Concl}}

We have applied a microscopic three-cluster model to investigate the nature of highly excited resonance states in $^{8}$Be. The model includes several three-cluster configurations, which are subsequently reduced to a large set of binary channels. One constituent of each binary channel is treated explicitly as a two-cluster subsystem. As a result, the model provides a more realistic description of the internal structure of interacting clusters, each possessing distinct clustering features. Moreover, this approach allows us to account for cluster polarization — i.e., the ability of clusters to alter their size and shape when approaching another cluster.

We performed a detailed analysis of the structure of the twin $1^+$, $2^+$, $3^+$, and $4^+$ resonance states and identified their dominant decay channels. In addition, we studied the $1^-$ and $2^-$ resonance states located above the $^7$Li+$p$ threshold. These investigations show that the present model provides a consistent and realistic description of the observed resonance spectrum near the $^7$Li+$p$ threshold. Furthermore, the results are in good agreement with those from other microscopic approaches, and generally offer a more accurate description of the twin resonance states.

We demonstrated that cluster polarization plays a critical role in the formation of $1^+$, $2^+$, $3^+$ and $4^+$ twin resonance states. In particular, we showed that with the chosen nucleon-nucleon potential, it is impossible to obtain two $2^{+}$ resonance states below the $p+^{7}$Li threshold without accounting for cluster polarization. Without polarization, one of these resonances appears slightly below the $p+^{7}$Li threshold, while the second remains above it. Inclusion of cluster polarization significantly lowers the second $2^{+}$ resonance state, placing both resonances in good agreement with experimental data.

In contrast, the energies of the $1^-$ and $2^-$ resonances near the $^7$Be+$n$ decay threshold of $^8$Be are largely insensitive to the degree of polarization of the binary subsystems. The $2^-$ resonance and the lowest $1^-$ state can be associated with neutron scattering on the ground and first excited states of $^7$Be, respectively, while the second $1^-$ resonance is consistent with a configuration of $^3$He scattering on $^5$He.

We proposed and implemented a simplified, yet physically motivated, two-channel model to study the formation mechanism of Feshbach-type $2^+$ resonances. This model includes the $p+^{7}$Li and $^{4}$He+$^{4}$He channels, which are characterized by a large difference in threshold energies. The coupling between these channels was scaled by a parameter $f_{C}$ ranging from 0 (uncoupled limit) to 1 (fully coupled case), allowing us to trace the evolution of the resonance spectrum. Although the physical model corresponds to full coupling ($f_{C} = 1$), the parametric variation of $f_C$ served as a diagnostic tool to reveal how near-threshold resonances emerge from bound states in the closed $p+^{7}$Li channel through coupling with the open $^{4}$He+$^{4}$He channel. We observed that these resonances evolve through a sequence of avoided crossings, where different closed-channel configurations successively dominate the near-threshold state. Despite this internal restructuring, the energy of the resonance remains nearly constant, indicating weak coupling to the open channel and a pronounced Feshbach character. Thus, our model shows that the $2^{+}$ Feshbach resonances in $^{8}$Be arise primarily from configuration mixing among closed channels and weak-to-moderate coupling with the continuum.
At the same time, a systematic channel-by-channel analysis demonstrated that additional binary channels—especially $^{3}$H+$^{5}$Li, $n+^{7}$Be, and $^{3}$He+$^{5}$He—play an essential role in lowering the energies of both $2^{+}$ resonances and reproducing their experimentally observed energy splitting, which cannot be achieved within the two-channel model alone.

As a future perspective of this work, in a forthcoming paper we plan to explore the dynamics of reactions induced by interactions of protons with $^{7}$Li, neutrons with $^{7}$Be, and deuterons with $^{6}$Li. The effects of resonance states determined in the present paper on cross sections and astrophysical $S$-factors of these reactions will be investigated in detail. Particular attention will be given to the low-energy region and reactions involving the production and destruction of the nuclei $^{6}$Li, $^{7}$Li, and $^{7}$Be. We anticipate that detailed insights into these reactions may contribute to resolving the cosmological lithium problem.

\begin{acknowledgments}

This work received partial support from the Program of Fundamental Research of the Physics and Astronomy Department of the National Academy of Sciences of Ukraine (Project No. 0122U000889). We extend our gratitude to the Simons Foundation for their financial support (Award ID: 1290598). Additionally, Y.L. acknowledges the National Institute for Nuclear Physics, Italy, for providing a research grant to support Ukrainian scientists.  
\end{acknowledgments}

\end{document}